\documentclass[aps,amsmath,amssymb,reprint,superscriptaddress,longbibliography,UTF8]{revtex4-1}

\usepackage{float}
\usepackage{graphicx}
\usepackage{dcolumn}
\usepackage{bm}
\usepackage{mathrsfs,amssymb}
\usepackage{ulem}
\usepackage[dvipsnames]{xcolor}
\usepackage[colorlinks=true,linkcolor=blue,citecolor=blue,anchorcolor=blue,urlcolor=blue]{hyperref}
\usepackage{lineno}

\expandafter\ifx\csname package@font\endcsname\relax\else
 \expandafter\expandafter
 \expandafter\usepackage
 \expandafter\expandafter
 \expandafter{\csname package@font\endcsname}%
\fi
\begin{document}

\title{Multiscale examination of strain effects in Nd--Fe--B permanent magnets}

\author{Min Yi}
\email{yi@mfm.tu-darmstadt.de}
\author{Hongbin Zhang}
\email{hzhang@tmm.tu-darmstadt.de}
\affiliation{Institute of Materials Science, Technische Universit\"{a}t Darmstadt, 64287 Darmstadt, Germany}
\author{Oliver Gutfleisch}
\email{gutfleisch@fm.tu-darmstadt.de}
\affiliation{Institute of Materials Science, Technische Universit\"{a}t Darmstadt, 64287 Darmstadt, Germany}
\author{Bai-Xiang Xu}
\email{xu@mfm.tu-darmstadt.de}
\affiliation{Institute of Materials Science, Technische Universit\"{a}t Darmstadt, 64287 Darmstadt, Germany}

\date{\today}

\begin{abstract}
We have performed a combined first-principles and micromagnetic study on the strain effects in Nd--Fe--B permanent magnets. First-principles calculations on Nd$_2$Fe$_{14}$B reveal that the magnetocrystalline anisotropy ($K$) is insensitive to the deformation along $c$ axis and the $ab$ in-plane shrinkage is responsible for the $K$ reduction. The predicted $K$ is more sensitive to the lattice deformation than what the previous phenomenological model suggests. The biaxial and triaxial stress states have a greater impact on $K$. Negative $K$ occurs in a much wider strain range in the $ab$ biaxial stress state. Micromagnetic simulations of Nd--Fe--B magnets using first-principles results show that a $3-4\%$ local strain in a 2-nm-wide region near the interface around the grain boundaries and triple junctions leads to a negative local $K$ and thus remarkably decreases the coercivity by $\sim60\%$ or $3-4$ T. The local $ab$ biaxial stress state is more likely to induce a large loss of coercivity. In addition to the local stress states and strain levels themselves, the shape of the interfaces and the intergranular phases also makes a difference. Smoothing the edge and reducing the sharp angle of the triple regions in Nd--Fe--B magnets would be favorable for a coercivity enhancement.
\end{abstract}

\maketitle

\section{Introduction}
Strain can be utilized to tailor the magnetic properties of many materials, leading to either promising applications or undesirable problems. For example, strain effects in soft magnetic materials can be used for the electric control of magnetic properties by using the strain mediated magnetoelectric coupling \cite{1zheng2004multiferroic,2wang2010multiferroic}. In addition, strain mediated magnetization switching has been a potential way to revolutionize the spintronic devices that currently utilize power-dissipating currents \cite{3yi2015effects,4tiercelin2016strain,5yi2015mechanically,6hu2016multiferroic,7yi2015180}. In the permanent magnets which are featured by high coercivity and high maximal energy product, local strain around the grain boundaries and triple junctions is thought to reduce the local magnetocrystalline anisotropy and thus the coercivity \cite{8woodcock2012understanding,9hrkac2014impact,10kubo2014development,11hrkac2014modeling,12hrkac2010role,13bance2014influence,14woodcock2014atomic}, degrading the magnetic performance. These indicate strain as a double-edged sword in magnetic materials. Understanding its effects is prerequisite for a wise application or avoidance of this double-edged sword.

In this work, we focus on the strain effects in a typical permanent magnet Nd--Fe--B. In Nd--Fe--B magnets, strain effects are inevitable. On one hand, sintering processes, post-thermal treatments, and hot pressing unavoidably induce a certain residual strain. Such strain can be either at the bulk level or at the local level. On the other hand, the coercivity of standard Nd--Fe--B magnets is only $\sim20\%$ of the theoretical upper limit from the Stoner--Wohlfarth model. The huge deviation from the theoretical prediction is believed to be mainly originated from the microstructural effects \cite{8woodcock2012understanding,15davies2001recent,16gutfleisch2011magnetic,new7sawatzki2016grain,new8loewe2015temperature}. The critical microstructural features that affect the coercivity are the intergrain phases and grain boundary phases. The structural or crystal-orientation mismatch between Nd$_2$Fe$_{14}$B main phase and other phases will generate local strain near the interfaces of different phases or grains. It is possible that such local strain results in regions of reduced anisotropy as nucleation sites for reversal domains.

For the theoretical study of strain effects in Nd--Fe--B magnets, by using the phenomenological theory regarding the magnetoelastic anisotropy \cite{17de1999magnetoelastic}, Hrkac et al. \cite{9hrkac2014impact,11hrkac2014modeling,12hrkac2010role,13bance2014influence,14woodcock2014atomic} and Kubo et al. \cite{10kubo2014development} used molecular dynamics (MD) to determine the strain induced anisotropy constant ($K_\text{me}$). However, depending on the interatomic potential used in MD, the value of calculated $K_\text{me}$ can differ in one order of magnitude. For example, based on a pairwise interaction model for Nd$_2$Fe$_{14}$B, Hrkac et al. considered various crystal structures and crystal orientations of Nd and Nd oxides and evaluated maximum values of $K_\text{me}\sim-10$−--$-4000$ MJ/m$^3$ in single atoms (average $K_\text{me}$ for all atoms: $\sim-1$−--$-10$ MJ/m$^3$) in a $\sim$2-nm local region \cite{9hrkac2014impact,11hrkac2014modeling,12hrkac2010role}. In contrast, Kubo et al. \cite{10kubo2014development} developed a new angular-dependent potential model for Nd$_2$Fe$_{14}$B and estimated $K_\text{me}$ in the order of $-0.1$ MJ/m$^3$ within a $\sim$2-nm region. However, no experimental results have directly verified this 2-nm local region with extremely reduced magnetocrystalline anisotropy. In fact, early experiments showed that the homogeneous thermal strain present at the boundaries of Nd$_2$Fe$_{14}$B grains has only a small influence on the coercivity \cite{17de1999magnetoelastic}. More recently, Murakami et al. \cite{18murakami2015strain,new10murakami2016strain} directly measured the strain distribution around different interfaces in sintered Nd--Fe--B magnets. They demonstrated that the region with a strain of $\varepsilon_c\sim\pm1\%$ was extended over several tens nanometer (not the theoretical prediction of $\sim$2 nm confined to a local region) away the interface. Similar to the early experiments \cite{17de1999magnetoelastic}, they also speculated that the interfacial strains have limited influence on the coercivity. One plausible reason for the inconsistence between simulations and experiments is the experimental resolution limitations, i.e. presently it is difficult to measure the strain within a $\sim$2-nm-wide local region in these experiments \cite{18murakami2015strain}. Therefore, in terms of the inconsistence not only between previous different MD simulations themselves but also between the simulations and experimental measurements up to now, in the modelling aspect it is highly required that this issue be more precisely investigated at a quantitative or multiscale level.

In the present work, we perform a combined first-principles and micromagnetic study on Nd--Fe--B magnets, in order to demonstrate a multiscale simulation framework for elucidating the strain effects on Nd--Fe--B magnets and to clarify what kind of local strain can significantly reduce the coercivity. Previous first-principles calculations have provided insights into the magnetic moments and the magnetocrystalline anisotropy based on either the crystal field of Nd ions \cite{19moriya2009first,20suzuki2014effects,21tanaka2011first,22tanaka2011first,23tanaka2011first,new11tatetsu2016first} or the total energy difference \cite{24drebov2013ab,25kitagawa2010magnetic}.
Especially, Suzuki et al. \cite{20suzuki2014effects} explored the crystal field parameter of Nd ions in the case of changing the length of $a$-axes and $c$-axis. Asali et al. \cite{new13asali2014dependence} showed the dependence of magnetic anisotropy on $c/a$ ratio of X$_2$Fe$_{14}$B (X=Y, Pr, Dy) and Torbatian et al. \cite{torbatian2014strain} examined triaxial-strain effects on the magnetic anisotropy in
Y$_2$Fe$_{14}$B. But they did not report results for Nd$_2$Fe$_{14}$B. So strain effects of Nd$_2$Fe$_{14}$B in different forms and magnitudes scrutinized from first principles are still of interests. By using the first-principles results as inputs, we carry out further micromagnetic simulations to elucidate the strain effects on the coercivity of single- and multi-grain Nd--Fe--B magnets.

\section{Methodology}
The first-principles calculations were carried in the framework of the projector augmented-wave formalism as implemented in the Vienna \textit{ab initio} simulation package (VASP) \cite{26kresse1996efficient}. The Perdew--Burke--Ernzerhof (PBE) exchange-correlation functional in the generalized gradient approximation (GGA) was employed \cite{new3perdew1996generalized}. According to the previous work \cite{24drebov2013ab}, an energy cutoff of 400 eV and a Monkhorst--Pack $k$--mesh $5\times5\times4$ were utilized to reach a good convergence. The convergence criteria for the full structure relaxation at different stress states and strain levels were set as 10$^{-5}$ eV and 10$^{-3}$ eV/\AA\;for the energies and forces, respectively \cite{24drebov2013ab}. To obtain the magnetocrystalline anisotropy ($K$), 4f electrons are treated as valance electrons \cite{24drebov2013ab}. Non-self-consistent calculations with different spin quantization axes were done by including spin-orbit coupling, starting from self-consistent charge densities of spin-polarized calculations. In this way, $K$ was evaluated as the change of such total energies when the magnetization was along different axes, i.e.
\begin{equation}
K=\begin{cases} \left[\text{max}\left(E_a, E_b\right)-E_c\right]/V & \left(E_a>E_c\;\text{and}\;E_b>E_c\right)\\
\left[\text{min}\left(E_a, E_b\right)-E_c\right]/V & \left(E_a<E_c\;\text{or}\;E_b<E_c\right)
\end{cases}
\end{equation}
in which $V$ is the volume of the relaxed unit cell. $E_a$, $E_b$, and $E_c$ are the total energy when the magnetization was parallel to $a$, $b$, and $c$ axis, respectively. Positive $K$ indicates easy axis along $c$ axis, while negative $K$ indicates an easy $ab$ plane.

Using $K$ and $M_\text{s}$ (saturation magnetization) obtained from first-principles calculations as functions of stress states and strain levels, micromagnetic simulations were carried out by the 3D NIST OOMMF code \cite{27oommf} for solving the Landau--Lifshitz--Gilbert (LLG) equation \cite{new4gilbert2004phenomenological,new5yi2014constraint,new9yi2016real}. Single- and multi-grain Nd--Fe--B magnets are discretized by cubic meshes with a size of 1 nm. For the single grain, prisms with different sectional geometry were considered. For modelling the multigrain, we used the scanning electron microscopy (SEM) image of a sintered Nd--Fe--B magnet from the previous experiments \cite{18murakami2015strain}. The exchange constant is set to be a constant of 12.5 pJ/m \cite{28sepehri2014micromagnetic}. Hysteresis curves were calculated by setting the initial magnetization along the positive $c$ axis and the external field along negative $c$ axis.

\begin{figure*}[!ht]
\centering
\includegraphics[width=9.5cm]{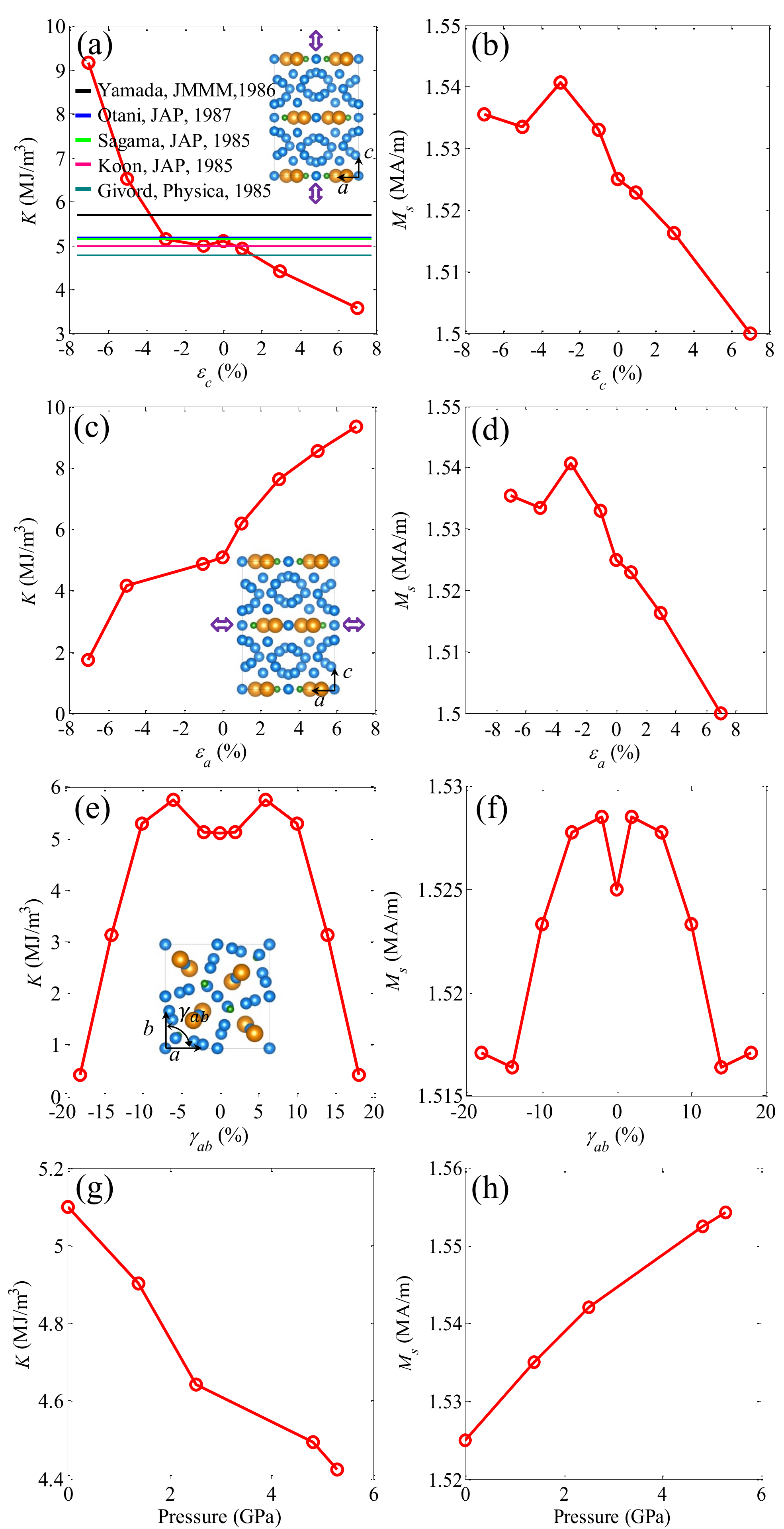}
\caption{First-principles calculated $K$ and $M_\text{s}$ in the stress state of (a) (b) $c$ uniaxial stress, (c) (d) $a$ uniaxial stress, (e) (f) pure shear in $ab$ plane, and (g) (h) hydrostatic pressure. The inset lines in (a) are corresponding to the experimental results of unstrained single crystal from the literatures \cite{29koon1985magnetic,30sagawa1985magnetic,31yamada1986magnetocrystalline,32otani1987magnetocrystalline,33givord1985structural}.}
\label{f1}
\end{figure*}

\begin{figure*}[t]
\centering
\includegraphics[width=16cm]{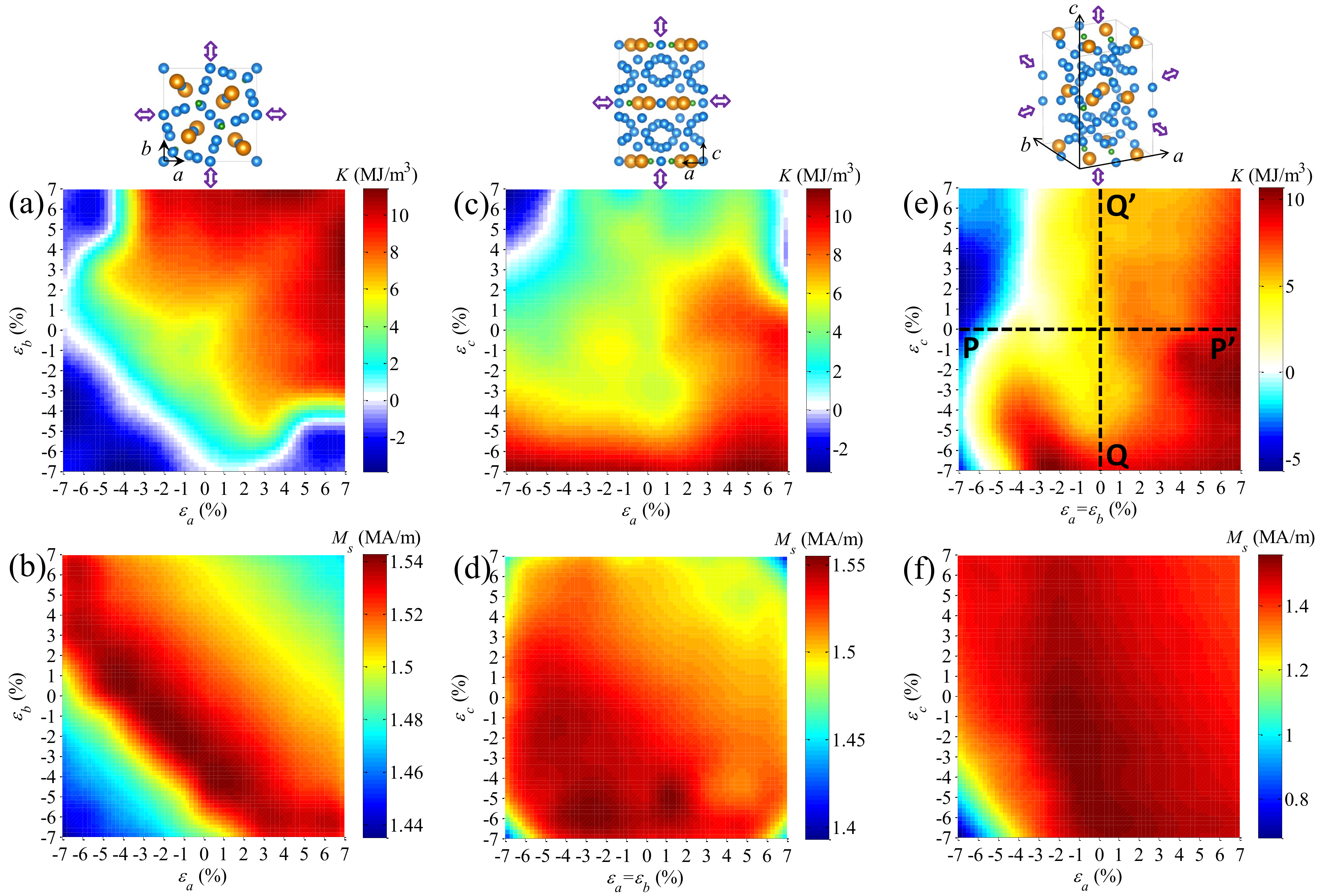}
\caption{First-principles calculated K and Ms in the stress state of (a) (b) $ab$ biaxial stress, (c) (d) $ac$ biaxial stress, and (e) (f) $abc$ triaxial stress.}
\label{f2}
\end{figure*}

\section{Results and discussion}
\subsection{First-principles calculations of strained Nd$_2$Fe$_{14}$B}
Figs. \ref{f1} and \ref{f2} present the first-principles results of $K$ and $M_\text{s}$ under various stress states and strain levels up to $\pm7\%$.
Bulk-level homogeneous strain in the above-mentioned range in real Nd$_2$Fe$_{14}$B magnet is not realistic. However, there are several plausible sources of the very large local strain, such as lattice or crystal orientation mismatch between Nd$_2$Fe$_{14}$B grains and the intergranular phases, thermal residual stress at triple regions, grain irregularity for stress concentration, symmetry breaking at the grain surface or near the interface, etc. In addition to these possible experimental phenomena, other reasons for introducing large strain range in this theoretical work lie in three aspects. Firstly, Hrkac et al. \cite{9hrkac2014impact,11hrkac2014modeling,12hrkac2010role,14woodcock2014atomic} and Kubo et al. \cite{10kubo2014development} adopted MD simulations and indeed predicted a significant change in $K$ that is caused by the very large local strain within a $\sim$2-nm narrow interface region. Secondly, up to now no experimental data on the strain in this localized region are available. The question on the magnitude of the extremely localized strain near the interface is still open from the experimental viewpoint. Thirdly, large strain is possible in the local region. It is well known in mechanics, theoretical strength of a material (inversely proportion to the square root of the atom layer distance) can be 3 orders higher than measurable fracture toughness (inversely proportion to the square root of the microcrack length) \cite{griffith1921phenomena}. Therefore the grain surface layer of 2 nm can sustain large deformation without fracture. 
Also as shown in the previous MD simulations \cite{9hrkac2014impact,11hrkac2014modeling,12hrkac2010role,14woodcock2014atomic,10kubo2014development}, the local atomic arrangements at the grain boundaries or interfaces experience dramatic change, but are still stable without fracture. The locally dramatic change in atomic arrangements can be considered as large effective strain which is confined to the vicinity of interfacial region. Our theoretical calculations are supposed as a first step to cover such a case in order to see what would occur there. 
Based on the above considerations, we introduced a large strain range (but still less than the strain values predicted by previous MD simulations) in this work. By inputting the strain-dependent first-principles results to the micromagnetic model with a locally strained region of $\sim$2 nm, we attempt to reveal the local strain effects on the coercivity. This theoretical work could be considered as a plausible first step towards a more accurate study by combining experimental local-strain measurement and theoretical calculations of a interface-containing large supercell with several hundreds of atoms. In the stress-free Nd$_2$Fe$_{14}$B unit cell, our first-principles calculations show a $K$ value of $\sim$5.1 MJ/m$^3$ ($\sim$30.1 meV/unit cell) and a $M_\text{s}$ value of $\sim$1.525 MA/m. The calculated $K$ agrees well with the experimental results \cite{29koon1985magnetic,30sagawa1985magnetic,31yamada1986magnetocrystalline,32otani1987magnetocrystalline,33givord1985structural}, as indicated by the five horizontal lines in Fig. \ref{f1}(a). The calculated $M_\text{s}$ is $\sim$38.3 $\mu_B$/formula unit (f.u.), which also matches well with the experimental value of $\sim$37.7 $\mu_B$/f.u. \cite{34herbst1991r} and other first-principle results \cite{35nordstrom1993calculation,36jaswal1990electronic,37kitagawa2009calculation}. The consistence between our calculations and experimental results validates our first-principles study on Nd$_2$Fe$_{14}$B.

\begin{figure*}[t]
\centering
\includegraphics[width=14.8cm]{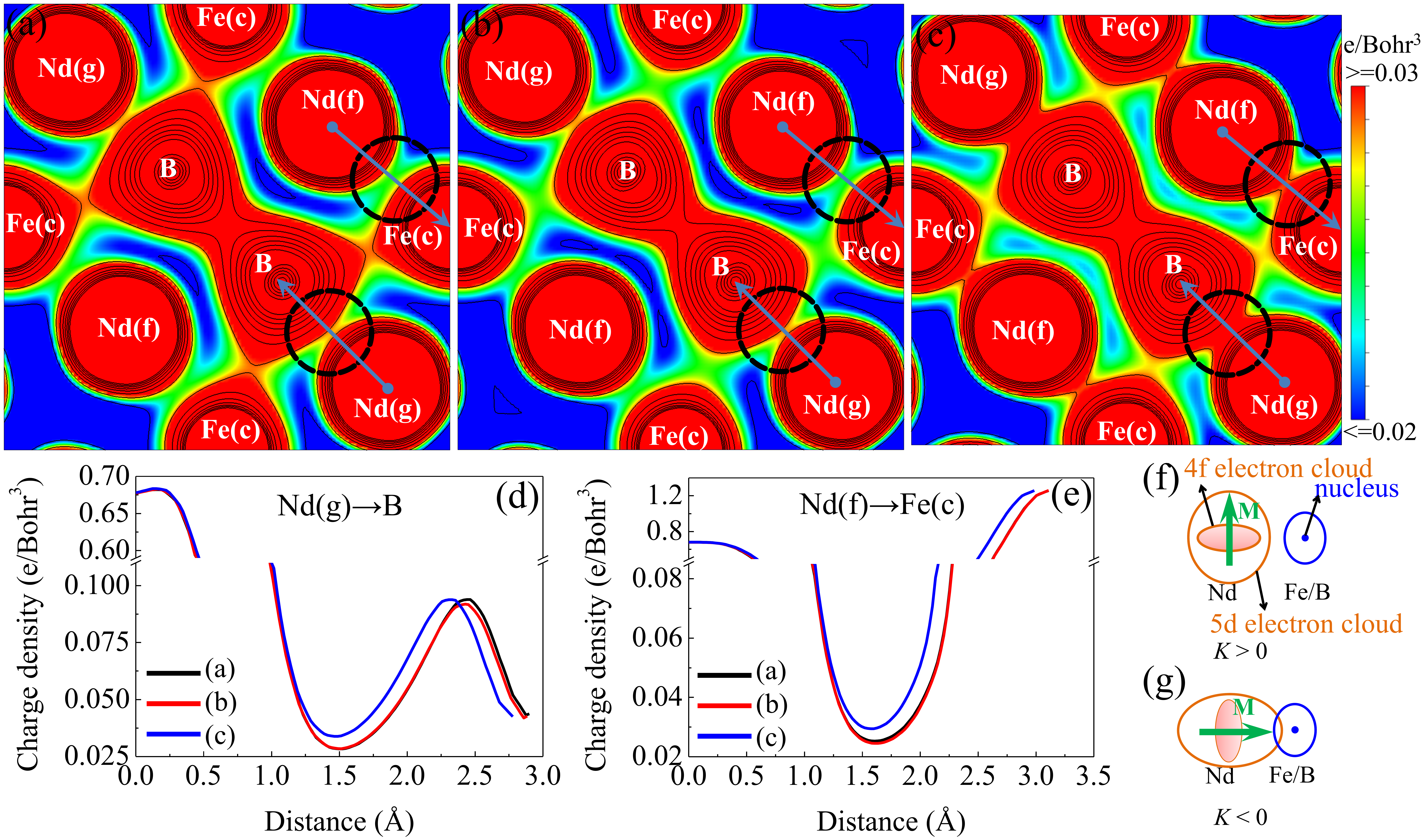}
\caption{Valence-electron-density distributions of Nd$_2$Fe$_{14}$B in (001) plane: (a) $\varepsilon_a=\varepsilon_b=\varepsilon_c=0$; (b) $\varepsilon_a=\varepsilon_b=0$ and $\varepsilon_c=4\%$; (c) $\varepsilon_a=\varepsilon_b=-4\%$ and $\varepsilon_c=4\%$.
Dotted circles indicate regions where charge density distribution around Nd atoms apparently changes. 
Charge density distribution along the lines (d) Nd(g)$\rightarrow$B and (e) Nd(f)$\rightarrow$Fe(c) indicated by arrows in (a)-(c). (f) and (g) Schematics for a possible explanation of the sign of $K$ \cite{22tanaka2011first}.}
\label{f3}
\end{figure*}

The calculated $M_\text{s}$ in Figs. \ref{f1} and \ref{f2} shows that it is not remarkably influenced by the stress states and strain levels, except the severer triaxial stress state in the lower left corner of Fig. \ref{f2}(f). However, the calculated $K$ is highly dependent both on the stress states and strain levels. In the $c$ uniaxial stress state in which only the crystal axis $c$ of Nd$_2$Fe$_{14}$B is stressed and other two crystal axes $a$ and $b$ are stress-free or free to relax, $K$ shows a decreasing trend as the strain $\varepsilon_c$ is increased, as shown in Fig. \ref{f1}(a). On the contrary, Fig. \ref{f1}(c) indicates that $K$ increases with the strain $\varepsilon_a$ when an $a$ uniaxial stress is applied. Fig. \ref{f1}(e) shows that the pure shear in the $ab$ plane has negligible effects on $K$ when the shear strain $\gamma_{ab}$ is less than $10\%$. Only in the extremely sheared case ($\gamma_{ab}>15\%$), $K$ is remarkably reduced.
Since the hydrostatic pressure up to $\sim$5.3 GPa induces a tiny shrinkage of the lattice, it only slightly reduces $K$, as shown in Fig. \ref{f1}(g) which is one special case of Fig. \ref{f2}(e). For the hydrostatic pressure in Fig. \ref{f1}(g), the stress in both three directions is the same. While for the triaxial stress state in Fig. \ref{f2}(e), the stress along $a(b)$ axis and the stress along $c$ axis can be either equal or not. The maximum hydrostatic pressure $\sim 5.3$ GPa in Fig. 1(g) corresponds to a strain state of $\varepsilon_a=\varepsilon_b \sim −1.27\%$ and $\varepsilon_c \sim −1.47\%$. The results in Fig. 1(g) are consistent with those in the triaxial stress state shown in Fig. 2(e).
The variation of $K$ under biaxial and triaxial stress states is presented in Fig. \ref{f2}. It is obvious that negative $K$ occurs in a much wider strain range in the $ab$ in-plane biaxial stress state, as shown in Fig. \ref{f2}(a). The shrinkage in $ab$ plane can notably reduce $K$. For example, an $ab$ biaxial stress state with $\varepsilon_a=\varepsilon_b=-3\%$ and $-4\%$ reduce $K$ to $\sim$1.4 and $\sim-$0.38 MJ/m$^3$, respectively. In contrast, for the $ac$ biaxial stress state in Fig. \ref{f2}(c), the strain range for negative $K$ is very small. Only in the case of negative $\varepsilon_a$ or large positive $\varepsilon_c$, $K$ is reduced. For the $abc$ triaxial stress state in Fig. \ref{f2}(e), negative $K$ appears for large negative $\varepsilon_a=\varepsilon_b$. The $c$ elongation and $ab$ plane shrinkage reduce $K$. For example, $K$ decreases to $\sim$2.1 MJ/m$^3$ in the case of $\varepsilon_a=\varepsilon_b=-3\%$ and $\varepsilon_c=3\%$. The results in Fig. \ref{f2}(e) agree well with the previous work by calculating the crystal field parameters of Nd ions \cite{20suzuki2014effects} and are qualitatively consistent with the results from MD simulations \cite{9hrkac2014impact,10kubo2014development,11hrkac2014modeling,12hrkac2010role}.

From the results in Fig. \ref{f1}(a) and (c), one might think that the shrinkage along $a$ and $c$ has opposite effects on $K$. In fact, this is not the case. Due to the positive Poisson effect, uniaxial tensile stress along $a$ ($c$) axis induces shrinkage along $c$ ($a$) axis. So under the uniaxial stress state (Fig. \ref{f1}(a) and (c)), the interference between the strain along $a$ and $c$ axis makes it difficult to judge the main influential factor for $K$. Then we consider the triaxial stress state in which we can either set strain along $c$ axis by forcing zero strains along $a$ and $b$, or set strain in $ab$ plane by forcing zero strains along $c$ axis, as indicated by the two dotted lines in Fig. \ref{f2}(e). For the line QQ$^\prime$, i.e. the case of $\varepsilon_a=\varepsilon_b=0$, $K$ does not change so much when $\varepsilon_c$ is larger than $-6\%$. For the line PP$^\prime$, i.e. the case of $\varepsilon_c=0$, $K$ gradually changes from $\sim$9.7 MJ/m$^3$ to $\sim-$4.5 MJ/m$^3$ when $\varepsilon_a=\varepsilon_b$ decreases from $7\%$ to $-7\%$. These results from lines PP$^\prime$ and QQ$^\prime$ indicate that $K$ is insensitive to the deformation along $c$, but changes apparently with the $ab$ in-plane deformation. In other words, the shrinkage in the $ab$ plane should be responsible for the $K$ reduction. The decrease of $K$ with increasing $\varepsilon_c$ in Fig. \ref{f1}(a) is ascribed to the $c$ elongation induced $ab$ plane shrinkage through the positive Poisson effect. These also explain the results in Fig. \ref{f2} that negative $K$ always appears in the region with negative $\varepsilon_a$ or $\varepsilon_b$ and positive $\varepsilon_c$ and $ab$ biaxial stress state allows much larger strain range for negative $K$.

In order to qualitatively understand the sign change of $K$, we analyzed the valence charge density. The density map in the (001) plane of Nd$_2$Fe$_{14}$B is shown for three typical cases in Fig. \ref{f3}. 
In order to clearly display the charge density difference, the legend is scaled down to the range $0.02-0.03$ e/Bohr$^3$. In this way, the charge density difference around Nd atomic sites can be easily identified by the color, as indicated by the dotted circles in Fig. \ref{f3}(a)-(c).
It can be found that the charge density at Fe(c) sites exhibits a distorted distribution towards B sites and forms an aspherical shape. The charge density at Nd(f) sites and Nd(g) sites is slightly different, but both deviate from the spherical distribution. The charge density at B sites is extremely anisotropic and is extended towards Nd(g) sites and Fe(c) sites \cite{new1gu1987comparative,new2ching1987electronic}. Despite of these common feature of the charge density, strain can induce some non-trivial changes. Comparison of Fig. \ref{f3}(a) with no strain and Fig. \ref{f3}(b) with $\varepsilon_a=\varepsilon_b=0$ and $\varepsilon_c=4\%$ reveals only a slight change of the charge distribution around Fe(c) and B sites. Since no remarkable change of the charge distribution around Nd sites is observed in Fig. \ref{f3}(b), the sign of $K$ remains the same as that in Fig. \ref{f3}(a). It indicates that deformation along $c$ axis without in-plane strain (i.e. $\varepsilon_a=\varepsilon_b=0$) does not remarkably change $K$, agreeing well with the above results.
In contrast, if an additional in-plane shrinkage strain ($\varepsilon_a=\varepsilon_b=-4\%$) is applied, the charge distribution around Nd sites is notably altered, as shown by the dotted circles in Fig. \ref{f3}(c). Moreover, charge density distribution along the lines Nd(g)$\rightarrow$B (Fig. \ref{f3}(d)) and Nd(f)$\rightarrow$Fe(c) (Fig. \ref{f3}(e)) also indicates apparent increase of charge density around Nd when in-plane compressive strain is applied.
Due to the reduction of the distance between Nd sites and Fe/B sites, there is evidence of some degree of hybridization between Fe/B atoms and Nd atoms (Fig. \ref{f3}(c)). Through the hybridization, the 5d electron cloud of Nd atoms apparently extends towards Fe/B atoms. This relocates the 4f electron cloud perpendicular to the $ab$ plane in order to avoid the repulsive force from the horizontally extended 5d electron cloud \cite{22tanaka2011first}, thus leading to an easy $ab$ plane and negative $K$ (Fig. \ref{f3}(g)). Therefore, one possible explanation is that the in-plane shrinkage makes Fe/B atoms much closer to Nd atoms and results in hybridization between them, which further changes the 5d electron cloud surrounding the 4f electron cloud of Nd atoms and finally alters the sign of $K$ \cite{22tanaka2011first}.

It should be noted that several researchers \cite{9hrkac2014impact,10kubo2014development,11hrkac2014modeling,12hrkac2010role} have dealt with the strain induced $K$ change by using the phenomenological magneto-elastic coupling energy which was derived by de Groot and de Kort \cite{17de1999magnetoelastic}. They calculated the strain induced anisotropy constant ($K_\text{me}$) as a function of lattice strain and applied $K_\text{me}$ to estimate the change of $K$ by using the elastic constants from isotropic polycrystals. For a qualitative and order-of-magnitude analysis, we rewrite the $K_\text{me}$ from de Groot and de Kort as $K_\text{me}\sim B\varepsilon$ in which $B$ denotes the magnetoelastic coefficient and $\varepsilon$ the strain level. By using the parameters given in the literature \cite{17de1999magnetoelastic}, our estimation of $B$ is shown to be in the order of 40 MJ/m$^3$. It means that a large strain in the order of $10\%$ can only give a $K$ change of $\sim4$ MJ/m$^3$. For a negative $K$, a strain more than $12\%$ is required. However, our first-principles calculations show that a small strain around $4\%$ can even reduce $K$ to negative values (Fig. \ref{f2}(a)). Hence our first-principles study indicates a much larger sensitivity of $K$ to the lattice deformation. The underestimation of strain effects by the phenomenological description could be attributed to the assumption of one-ion magneto-elastic Hamiltonian without the two-ion one, because the two-ion magneto-elasticity is also related to the modification of the two-ion magnetic interactions by the strains \cite{38morin1990quadrupolar}. But in our first-principles calculations, both one-ion and two-ion magnetic interactions, as well as the fully electron-lattice coupling, are consistently included.

\subsection{Micromagnetic simulations of locally strained Nd--Fe--B magnets}
Previous experiments have demonstrated that homogeneous small strain in Nd--Fe--B magnets has negligible effect on the coercivity \cite{17de1999magnetoelastic}. However, previous MD simulations verified that a large strain is possible in a very localized $\sim$2-nm-wide region near the interface \cite{9hrkac2014impact,10kubo2014development,11hrkac2014modeling,12hrkac2010role}. They used the atomic displacement near the interface to calculate the local strain, which is taken as the lattice strain as inputs for the phenomenological magneto-elastic theory \cite{17de1999magnetoelastic} to estimate the $K$ change. In the micromagnetic simulations here, we also follow the similar idea as shown in these previous studies  \cite{9hrkac2014impact,11hrkac2014modeling,12hrkac2010role,14woodcock2014atomic,10kubo2014development,new12zickler2015nanoanalytical}, i.e. the source of the local strain is not the focus and an effective lattice strain is assigned to the local region. The symmetry breaking and the change of chemical environments near the local region are out of the scope in this work, although they can also influence the coercivity. However, unlike these previous studies which used the phenomenological theory \cite{17de1999magnetoelastic}, here we directly take the lattice strains and stress states associated with the first-principles calculations to define the locally strained region in Nd--Fe--B magnets. The local region is approximately set as 2 nm thick, as demonstrated by the MD simulations \cite{9hrkac2014impact,10kubo2014development,11hrkac2014modeling,12hrkac2010role,14woodcock2014atomic}. The parameters $K$ and $M_\text{s}$ of the locally strained region under various strain levels and stress states are taken from the first-principles results presented above.

\begin{figure}[!hb]
\centering
\includegraphics[width=8.6cm]{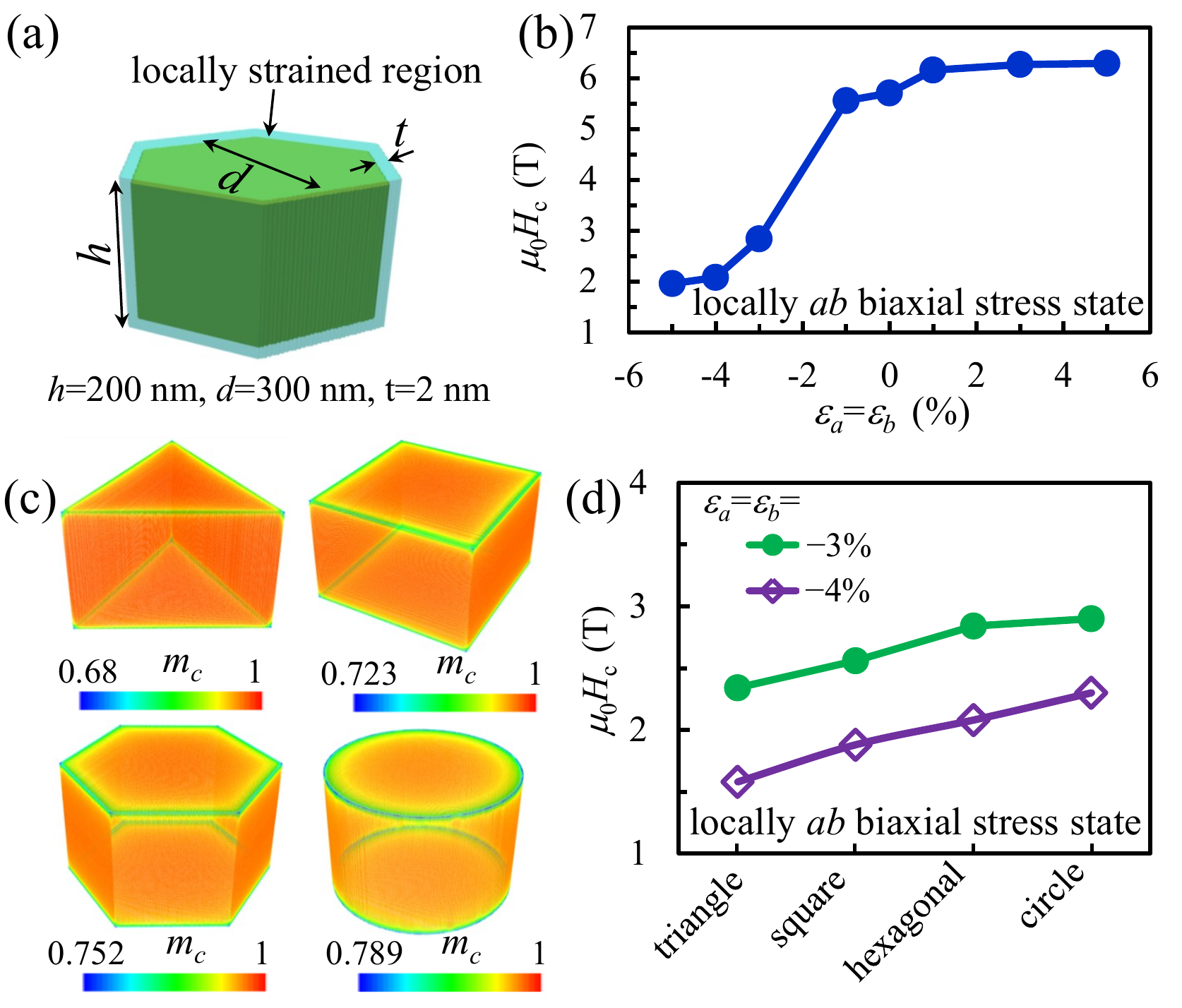}
\caption{(a) Schematic of a single-grain Nd--Fe--B magnet with a hexagonal section and with its surface covered by a locally strained region. 
(b) Local strain dependent coercivity for the grain in (a) under the local $ab$ biaxial stress state.
(c) $m_c$ distribution at the remanent state ($\mu_0H_\text{ex}=0$) for grains with triangular, square, hexagonal, and circular sections under a local $ab$ biaxial strain of $-4\%$.
(d) Coercivity for the grains in (c) under local $ab$ biaxial strains of $-3\%$ and $-4\%$.}
\label{f4}
\end{figure}

\begin{figure*}[!ht]
\centering
\includegraphics[width=17cm]{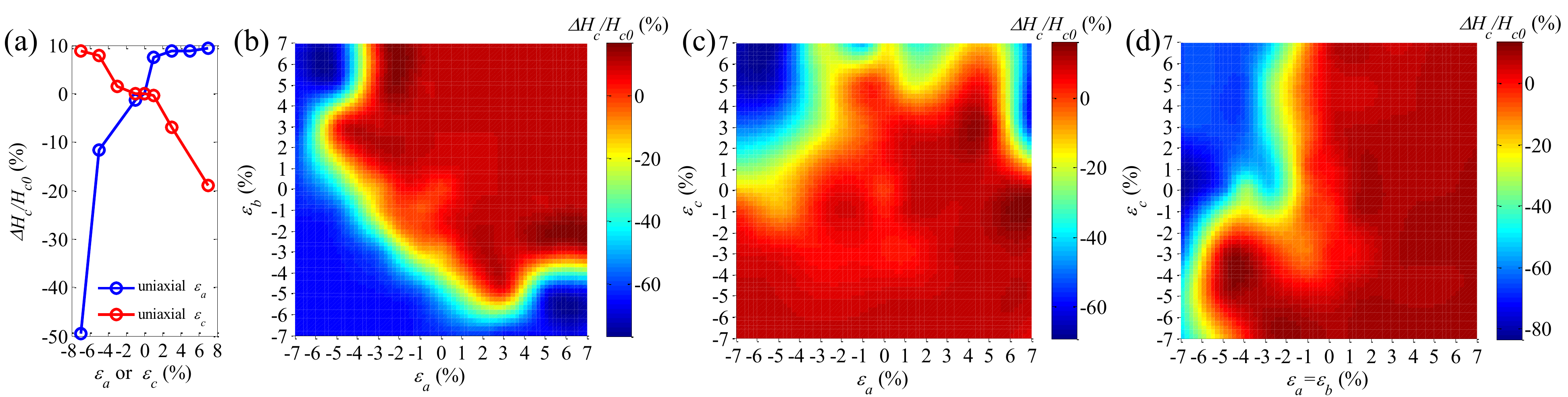}
\caption{Micromagnetic simulation results on the coercivity change as functions of the local surface strain under the stress state of (a) uniaxial stress, (b) $ab$ biaxial stress, (c) $ac$ biaxial stress, and (d) $abc$ triaxial stress. The micromagnetic mode is single Nd--Fe--B grain with a hexagonal section in Fig. \ref{f4}(a).}
\label{f5}
\end{figure*}

\subsubsection{Single-grain Nd--Fe--B magnets}
We firstly investigated the prism-shaped single grain which is covered by a locally strained surface with a thickness of $t$. Fig. \ref{f4}(a) displays the grain shape of a hexagonal prism, with the geometry dimension of $h=200$ nm, $d=300$ nm, and $t=2$ nm. If we assume that the grain surface is under the local $ab$ biaxial stress state, it can be found that for the hexagonal prism, the coercivity decreases from 5.7 T to 1.96 T under an $ab$ biaxial strain of $\varepsilon_a=\varepsilon_b=-5\%$ (\ref{f4}(b)). However, the coercivity is only slightly increased by 0.5 T in the case of $\varepsilon_a=\varepsilon_b=5\%$. This indicates that the coercivity is more sensitive to the local region with $ab$ plane shrinkage and negative $K$.

We further studied the effects of grain shape of the locally strained single grain. 
The motivation is to explore the possible role of the place where strain/stress appears and the associated micromagnetic mechanism.
The grain shape effects have been recently investigated for achieving high coercivity \cite{39yi2016micromagnetic,40erokhin2016optimization,41bance2014micromagnetics,42bance2014grain}. Here we considered four types of prism grains with triangular, rectangular, hexagonal, and circular sections. The distribution of the $c$ component of the unit magnetization vector ($m_c$) at the remanent state ($\mu_0H_\text{ex}=0$) is presented in Fig. \ref{f4}(c). It can be seen that the magnetization near the corners or edges has already rotated out of the easy direction even at the remanent state. More precisely, the minimum $m_c$ values in Fig. \ref{f4}(c) are found to decrease in the order: circular prism $>$ hexagonal prism $>$ square prism $>$ triangular prism. This means that the local reversal occurs fastest in the triangular prism and slowest in the circular prism. Such a local reversal is due to the inhomogeneous stray field near the corners or edges in the nonellipsoidal grains \cite{39yi2016micromagnetic,43schrefl1994nucleation}. By the local reversal, the inhomogeneous magnetization can suppress magnetic surface charges and decrease the stray-field energy with respect to the homogeneous magnetic state. The different local reversal behavior could result in distinct coercivity. We find in Fig. \ref{f4}(d) that at the same local stress states and strain levels, the coercivity is shown to increase in the order: triangular prism $<$ square prism $<$ hexagonal prism $<$ circular prism. For example, in the case of a local $ab$ biaxial stress state with $\varepsilon_a=\varepsilon_b=-4\%$, the coercivity is found to significantly increase from 1.58 T in the triangular prism to 2.3 T in the circular prism. These results indicate that in addition to the local stress states and strain levels themselves, where the locally strained region appears (e.g. grain shape, surface irregularity, edge curvature, etc.) also plays an important role in determining the coercivity.

By using the hexagonal prism in Fig. \ref{f4}(a), we carried out a detailed study on the sensitivity of the coercivity to the local stress states and strain levels. The coercivity change distribution in Fig. \ref{f5} is similar to the $K$ distribution in Figs. \ref{f1} and \ref{f2}. We find that in all the cases, the coercivity enhancement is limited to $\sim10\%$, while the coercivity decrease can be as high as $\sim80\%$. Again, the coercivity decrease is found to be more sensitive to the local strain than the coercivity enhancement. Fig. \ref{f5}(a) shows that the uniaxial stress state along $c$ axis and $a$ axis induce a maximum coercivity decrease of $\sim20\%$ and $\sim50\%$, respectively. The local $ab$ biaxial stress state allows a much larger strain range for the coercivity decrease by $\sim60\%$, as shown in Fig. \ref{f5}(b). In contrast, the local $ac$ biaxial and $abc$ triaxial stress states have much smaller strain range for the coercivity decrease, as shown in Fig. \ref{f5}(c) and (d).

\subsubsection{Multi-grain Nd--Fe--B magnets}
\begin{figure}[!t]
\centering
\includegraphics[width=8.3cm]{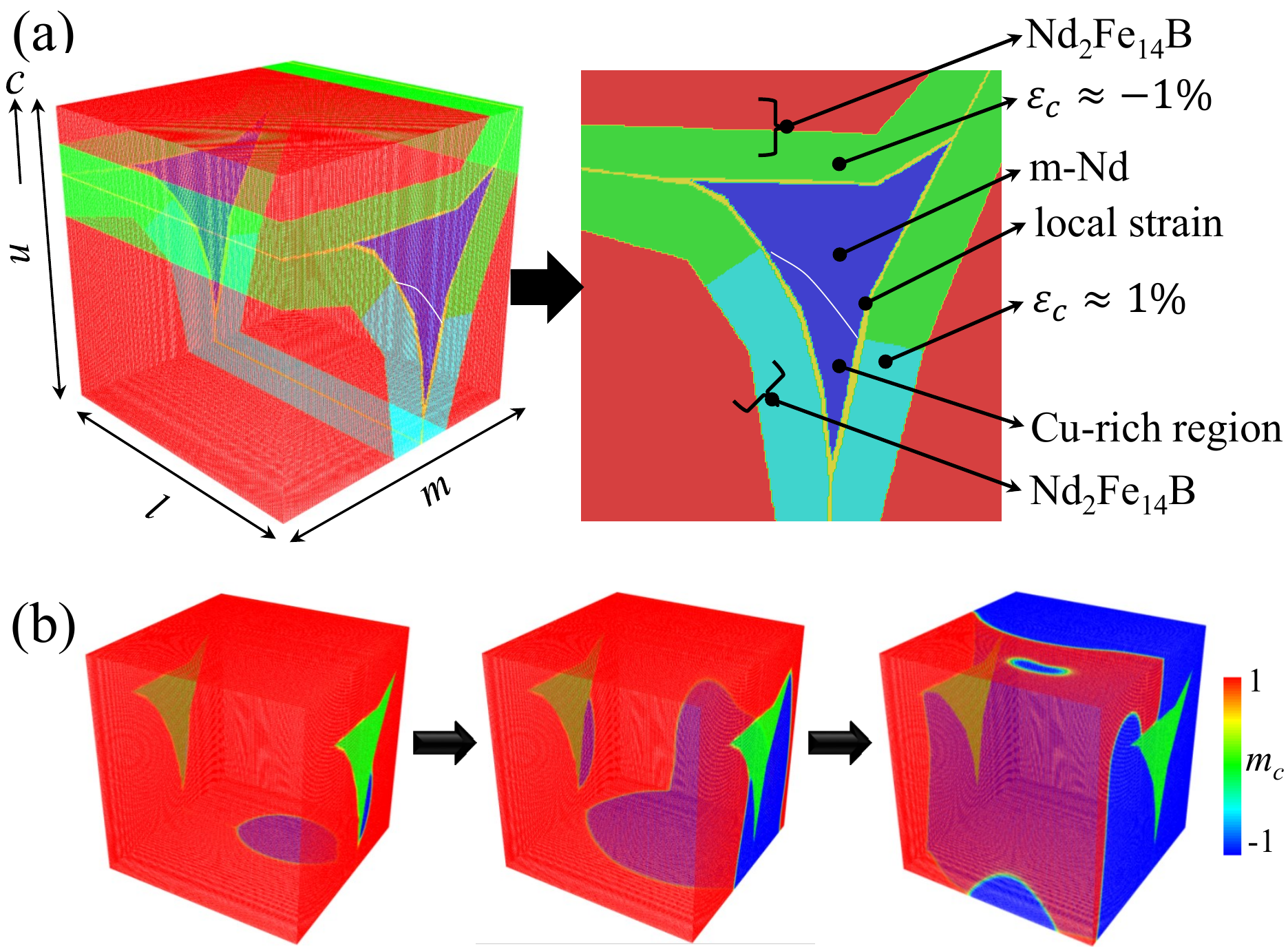}
\caption{Micromagnetic simulation of a multigrain based on the SEM image of the experimental work \cite{18murakami2015strain}. (a) Model illustration with the size of $l=m=280$ nm and $n=300$ nm. The locally strained region is with a width of $\sim$2 nm. (b) Magnetic reversal process between external fields of $-1.96$ T and $-1.98$ T when the local region is under the biaxial stress state with $\varepsilon_a=\varepsilon_b=-4\%$.}
\label{f6}
\end{figure}

\begin{figure}[!t]
\centering
\includegraphics[width=7cm]{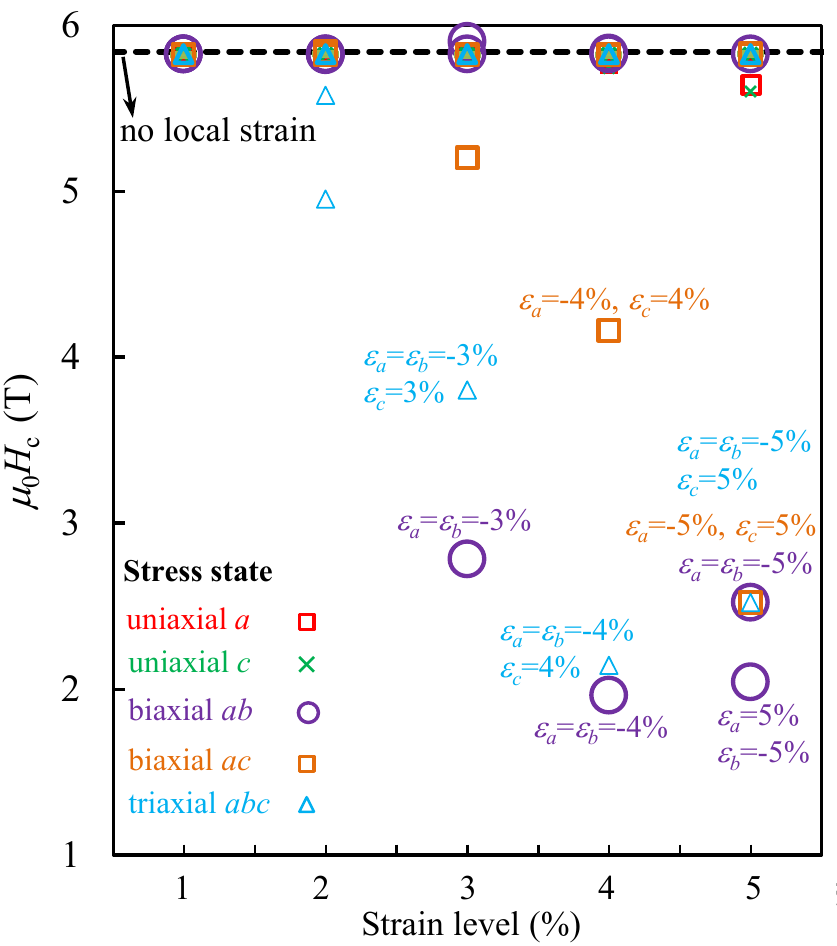}
\caption{Coercivity of the multigrain as functions of the local stress states and strain levels. The dotted line indicates the coercivity in the case of no local strain. Each color denotes one stress state of the local region. The strain values are marked when the coercivity decrease is over 1 T.}
\label{f7}
\end{figure}

Micromagnetic simulations on the multigrain were further performed. The multigrain model in Fig. \ref{f6}(a) was built by using the SEM image of a sintered Nd--Fe--B magnet \cite{18murakami2015strain}. The size $m=280$ nm and $n=300$ nm is estimated from the SEM image. Around the triple junction, the region with a strain of $\varepsilon_c=\pm1\%$ is extended over several tens nanometer away the interface, as measured in the previous experimental work \cite{18murakami2015strain}. An additional locally strained region with 2-nm width is assumed in the interface, as did in the previous work \citep{11hrkac2014modeling,9hrkac2014impact,12hrkac2010role}.
 Due to the small size of the model, the simulated coercivity without local strain is as high as $\sim$5.84 T, as shown in Fig. \ref{f7}. The effect of the strain $\varepsilon_c=\pm 1\%$ extending over several tens nanometer on the coercivity is neglectable, further confirming the statement in the previous experimental work \cite{18murakami2015strain}. It can be seen from Fig. \ref{f7} that in the case of local uniaxial stress states, the strain has little influence on the coercivity. However, the biaxial and triaxial stress states remarkably reduce the coercivity. An $ab$ biaxial stress state with $\varepsilon_a=\varepsilon_b=-4\%$ and $-3\%$ decreases the coercivity from 5.84 T to 1.98 T and 2.8 T or by  $\sim 66\%$ and $\sim 52\%$, respectively. This means that a moderate strain level in a suitable local stress state can reduce the coercivity in the multigrain Nd--Fe--B magnets by more than 3 T. 
The magnetic reversal process in Fig. \ref{f6}(b) indicates several apparent nucleation sites in the locally strained region. Then the reversal domain rapidly expands and the whole grain is completely reversed instantly.

\begin{figure}[!h]
\centering
\includegraphics[width=7cm]{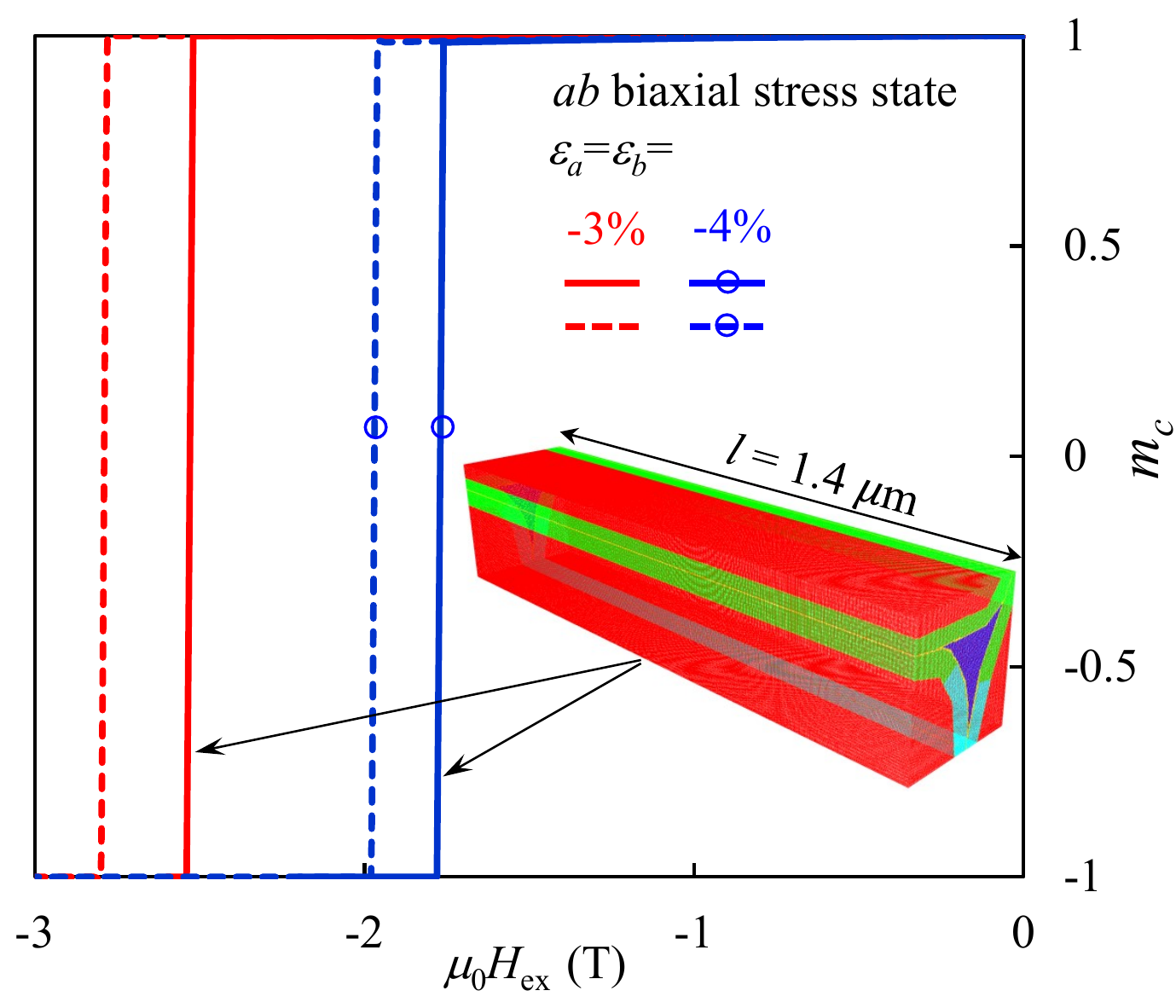}
\caption{Simulated reversal curves of the multigrain with the local strain region under the $ab$ biaxial stress state ($\varepsilon_a=\varepsilon_b=-3\%$ and $-4\%$). Solid lines correspond to the model in the inset with $l=1.4$ $\mu$m. Dotted lines correspond to the model in Fig. \ref{f6}(a).}
\label{f8}
\end{figure}

\begin{figure}[!h]
\centering
\includegraphics[width=7cm]{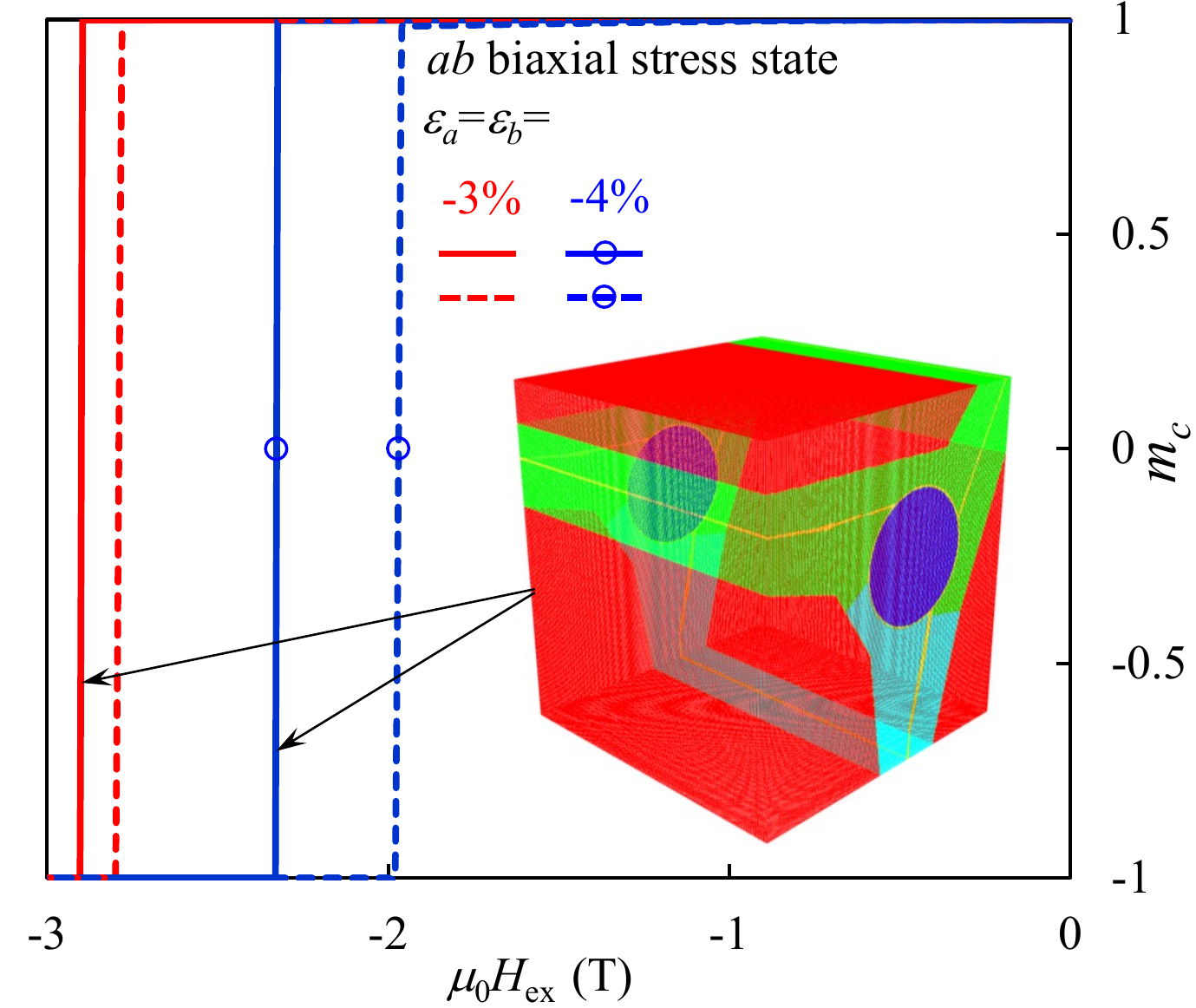}
\caption{Simulated reversal curves of the multigrain with the local strain region under the $ab$ biaxial stress state ($\varepsilon_a=\varepsilon_b=-3\%$ and $-4\%$). Solid lines correspond to the model in the inset with a circularly shaped triple region. Dotted lines correspond to the model in Fig. \ref{f6}(a).}
\label{f9}
\end{figure}

By using the first-principles results in Figs. \ref{f1} and \ref{f2}, we calculated the coercivity of the multigrain as functions of the local stress states and strain levels, as shown in Fig. \ref{f7}. It is found that the local uniaxial stress states almost do not affect the coercivity. Only the strain values marked in Fig. \ref{f7} and their associated stress states can reduce the coercivity by more than 1 T. Obviously, the local $ab$ biaxial stress state possesses a higher possibility to result in more reduction in the coercivity. In the case of local $ab$ biaxial and $abc$ triaxial stress in Fig. \ref{f7}, though $5\%$ strain induce a large negative value of $K$ than $4\%$ strain, the coercivity is slightly increased. This is due to the fact that a local negative $K$ favors the formation of an initial 90 degree domain wall between the locally strained region and the strain-free region, but a more negative $K$ increases the field for the subsequent formation of a 180 degree domain wall. This also indicates that a larger negative value of local $K$ inducing more reduction in the coercivity is not always correct.

Due to the small mesh size (1 nm here) determined by the physical length in micromagnetic simulations, the sample size of the modeled multigrain is often much smaller than that of the real magnets. In order to present an example for demonstrating the size effect, here we increase the multigrain size by extending $l$ (Fig. \ref{f6}(a)) to 1.4 $\mu$m, as show in the inset of Fig. \ref{f8}. Such an extension results in a mesh number of $\sim$0.12 billion, which is extremely computationally expensive. From the reversal curves in Fig. \ref{f8}, it can be seen that in both local $ab$ biaxial stress states with $\varepsilon_a=\varepsilon_b=-4\%$ and $-3\%$, the increase of $l$ from 280 nm to 1.4 $\mu$m makes the coercivity decrease by $\sim$ 0.3 T. The coercivity reduction can be qualitatively understood from the demagnetization effect. The $l$ extension favors a reduction and increase of demagnetization factor along the $l$ and $c$ direction, respectively. This increases the in-plane shape anisotropy and exerts additional torque to make magnetization deviate from $c$ axis, thus resulting in premature nucleation and reduced coercivity.

Finally it is worth mentioning that the single-grain results in Fig. \ref{f5}(d) inspire a strategy of increasing the coercivity by tuning the shape or geometry of the local strain region in the multigrains. By this inspiration, we then changed the model in Fig. \ref{f6}(a) into an ideal model whose triple region is set as a circular prism, as shown in the inset of Fig. \ref{f9}, in order to study the effect of where the local strain appears. In contrast to Fig. \ref{f6}(a) in which the local strain is in the triangular edge of the triple region, the model in Fig. \ref{f9} puts the local strain in the circular edge. We can find from Fig. \ref{f9} that the coercivity is obviously enhanced if a circular prism is used to represent the triple region. This is consistent with the result from the single-grain study and indicates the shape of the triple region as an influential factor for the coercivity. Smoothing the edge and removing the sharp angle of the triple region are favorable for the coercivity enhancement.
It should be noted that the circularly shaped triple region presented in the simulation is an ideal case, but it provide practical information towards coercivity enhancement. In realistic condition, though achieving a perfectly circular triple region is difficult, reducing the sharp angle of the triple regions or making them as smooth as possible in Nd--Fe--B magnets is possible by using gas atomised powders, controlled grain boundary diffusion, or additive manufacturing.

\section{Conclusions}
The strain effects in Nd--Fe--B magnets are examined by a combined first-principles and micromagnetic study. In this way, we use the first-principles results on the stress states and strain levels dependent $K$ and $M_\text{s}$ as the input for micromagnetic simulations of the coercivity in single- and multi-grain Nd--Fe--B magnets. The main conclusions are summarized in the following:

(1) In Nd$_2$Fe$_{14}$B phase, the stress states and strain levels have negligible effects on $M_\text{s}$ but significant effects on $K$. $K$ is sensitive to the $ab$ in-plane deformation rather than the $c$-axis deformation. The $ab$ plane shrinkage is responsible for the $K$ reduction. The biaxial and triaxial stress states have a greater impact on $K$ than other stress states. Negative $K$ occurs in a much wider strain range in the $ab$ biaxial stress state.

(2) $K$ is shown to be more sensitive to the lattice deformation by the first-principles study than by the previous phenomenological model \cite{17de1999magnetoelastic} which only considers one-ion magneto-elastic Hamiltonian and underestimates the strain effects. An $ab$ biaxial stress state with $\varepsilon_a=\varepsilon_b=-3\%$ and $-4\%$ reduces $K$ to $\sim$1.4 and $\sim-0.38$ MJ/m$^3$, respectively.

(3) In Nd--Fe--B magnets, the local $ab$ biaxial stress state in the locally strained region is more likely to induce a large loss of coercivity. A coercivity decrease by $60\%$ or by $3-4$ T can be induced by a $3-4\%$ local strain in a 2-nm-wide region near the interface around the grain boundaries and triple junctions.

(4) In addition to the local stress states and strain levels themselves, the shape of the interfaces and the intergranular phases also makes a difference. Smoothing the edge and reducing the sharp angle of the triple regions in Nd--Fe--B magnets would be favorable for a coercivity enhancement.

It is anticipated that our multiscale results here based on first-principles calculations and micromagnetic simulations provide quantitative information for that what kind of local stress state and how large local strain can induce significant decrease in the coercivity. The results will also be applied to the high-resolution experimental research of the local strain measurement in Nd--Fe--B permanent magnets.

\section*{Acknowledgment}
The financial supports from the German federal state of Hessen through its excellence programme LOEWE “RESPONSE” and the German Science Foundation (DFG Xu121/7-1) are appreciated. The authors also greatly acknowledge the access to the Lichtenberg High Performance Computer of TU Darmstadt.

\bibliography{mybibfile}

\begin{thebibliography}{57}%
\makeatletter
\providecommand \@ifxundefined [1]{%
 \@ifx{#1\undefined}
}%
\providecommand \@ifnum [1]{%
 \ifnum #1\expandafter \@firstoftwo
 \else \expandafter \@secondoftwo
 \fi
}%
\providecommand \@ifx [1]{%
 \ifx #1\expandafter \@firstoftwo
 \else \expandafter \@secondoftwo
 \fi
}%
\providecommand \natexlab [1]{#1}%
\providecommand \enquote  [1]{``#1''}%
\providecommand \bibnamefont  [1]{#1}%
\providecommand \bibfnamefont [1]{#1}%
\providecommand \citenamefont [1]{#1}%
\providecommand \href@noop [0]{\@secondoftwo}%
\providecommand \href [0]{\begingroup \@sanitize@url \@href}%
\providecommand \@href[1]{\@@startlink{#1}\@@href}%
\providecommand \@@href[1]{\endgroup#1\@@endlink}%
\providecommand \@sanitize@url [0]{\catcode `\\12\catcode `\$12\catcode
  `\&12\catcode `\#12\catcode `\^12\catcode `\_12\catcode `\%12\relax}%
\providecommand \@@startlink[1]{}%
\providecommand \@@endlink[0]{}%
\providecommand \url  [0]{\begingroup\@sanitize@url \@url }%
\providecommand \@url [1]{\endgroup\@href {#1}{\urlprefix }}%
\providecommand \urlprefix  [0]{URL }%
\providecommand \Eprint [0]{\href }%
\providecommand \doibase [0]{http://dx.doi.org/}%
\providecommand \selectlanguage [0]{\@gobble}%
\providecommand \bibinfo  [0]{\@secondoftwo}%
\providecommand \bibfield  [0]{\@secondoftwo}%
\providecommand \translation [1]{[#1]}%
\providecommand \BibitemOpen [0]{}%
\providecommand \bibitemStop [0]{}%
\providecommand \bibitemNoStop [0]{.\EOS\space}%
\providecommand \EOS [0]{\spacefactor3000\relax}%
\providecommand \BibitemShut  [1]{\csname bibitem#1\endcsname}%
\let\auto@bib@innerbib\@empty
\bibitem [{\citenamefont {Zheng}\ \emph {et~al.}(2004)\citenamefont {Zheng},
  \citenamefont {Wang}, \citenamefont {Lofland}, \citenamefont {Ma},
  \citenamefont {Mohaddes-Ardabili}, \citenamefont {Zhao}, \citenamefont
  {Salamanca-Riba}, \citenamefont {Shinde}, \citenamefont {Ogale},
  \citenamefont {Bai}, \citenamefont {Viehland}, \citenamefont {Jia},
  \citenamefont {Schlom}, \citenamefont {Wuttig}, \citenamefont {Roytburd},\
  and\ \citenamefont {Ramesh}}]{1zheng2004multiferroic}%
  \BibitemOpen
  \bibfield  {author} {\bibinfo {author} {\bibfnamefont {H.}~\bibnamefont
  {Zheng}}, \bibinfo {author} {\bibfnamefont {J.}~\bibnamefont {Wang}},
  \bibinfo {author} {\bibfnamefont {S.~E.}\ \bibnamefont {Lofland}}, \bibinfo
  {author} {\bibfnamefont {Z.}~\bibnamefont {Ma}}, \bibinfo {author}
  {\bibfnamefont {L.}~\bibnamefont {Mohaddes-Ardabili}}, \bibinfo {author}
  {\bibfnamefont {T.}~\bibnamefont {Zhao}}, \bibinfo {author} {\bibfnamefont
  {L.}~\bibnamefont {Salamanca-Riba}}, \bibinfo {author} {\bibfnamefont
  {S.~R.}\ \bibnamefont {Shinde}}, \bibinfo {author} {\bibfnamefont {S.~B.}\
  \bibnamefont {Ogale}}, \bibinfo {author} {\bibfnamefont {F.}~\bibnamefont
  {Bai}}, \bibinfo {author} {\bibfnamefont {D.}~\bibnamefont {Viehland}},
  \bibinfo {author} {\bibfnamefont {Y.}~\bibnamefont {Jia}}, \bibinfo {author}
  {\bibfnamefont {D.~G.}\ \bibnamefont {Schlom}}, \bibinfo {author}
  {\bibfnamefont {M}~\bibnamefont {Wuttig}}, \bibinfo {author} {\bibfnamefont
  {A}~\bibnamefont {Roytburd}}, \ and\ \bibinfo {author} {\bibfnamefont
  {R}~\bibnamefont {Ramesh}},\ }\bibfield  {title} {\enquote {\bibinfo {title}
  {Multiferroic {BaTiO$_3$--CoFe$_2$O$_4$} nanostructures},}\ }\href {\doibase
  10.1126/science.1094207} {\bibfield  {journal} {\bibinfo  {journal}
  {Science}\ }\textbf {\bibinfo {volume} {303}},\ \bibinfo {pages} {661--663}
  (\bibinfo {year} {2004})}\BibitemShut {NoStop}%
\bibitem [{\citenamefont {Wang}\ \emph {et~al.}(2010)\citenamefont {Wang},
  \citenamefont {Hu}, \citenamefont {Lin},\ and\ \citenamefont
  {Nan}}]{2wang2010multiferroic}%
  \BibitemOpen
  \bibfield  {author} {\bibinfo {author} {\bibfnamefont {Y.}~\bibnamefont
  {Wang}}, \bibinfo {author} {\bibfnamefont {J.}~\bibnamefont {Hu}}, \bibinfo
  {author} {\bibfnamefont {Y.}~\bibnamefont {Lin}}, \ and\ \bibinfo {author}
  {\bibfnamefont {C.-W.}\ \bibnamefont {Nan}},\ }\bibfield  {title} {\enquote
  {\bibinfo {title} {Multiferroic magnetoelectric composite nanostructures},}\
  }\href {\doibase 10.1038/asiamat.2010.32} {\bibfield  {journal} {\bibinfo
  {journal} {NPG Asia Mater.}\ }\textbf {\bibinfo {volume} {2}},\ \bibinfo
  {pages} {61--68} (\bibinfo {year} {2010})}\BibitemShut {NoStop}%
\bibitem [{\citenamefont {Yi}\ \emph {et~al.}(2015{\natexlab{a}})\citenamefont
  {Yi}, \citenamefont {Xu},\ and\ \citenamefont {Shen}}]{3yi2015effects}%
  \BibitemOpen
  \bibfield  {author} {\bibinfo {author} {\bibfnamefont {M.}~\bibnamefont
  {Yi}}, \bibinfo {author} {\bibfnamefont {B.-X.}\ \bibnamefont {Xu}}, \ and\
  \bibinfo {author} {\bibfnamefont {Z.}~\bibnamefont {Shen}},\ }\bibfield
  {title} {\enquote {\bibinfo {title} {Effects of magnetocrystalline anisotropy
  and magnetization saturation on the mechanically induced switching in
  nanomagnets},}\ }\href {\doibase 10.1063/1.4914485} {\bibfield  {journal}
  {\bibinfo  {journal} {J. Appl. Phys.}\ }\textbf {\bibinfo {volume} {117}},\
  \bibinfo {pages} {103905} (\bibinfo {year} {2015}{\natexlab{a}})}\BibitemShut
  {NoStop}%
\bibitem [{\citenamefont {Tiercelin}\ \emph {et~al.}(2016)\citenamefont
  {Tiercelin}, \citenamefont {Dusch}, \citenamefont {Giordano}, \citenamefont
  {Klimov}, \citenamefont {Preobrazhensky},\ and\ \citenamefont
  {Pernod}}]{4tiercelin2016strain}%
  \BibitemOpen
  \bibfield  {author} {\bibinfo {author} {\bibfnamefont {N.}~\bibnamefont
  {Tiercelin}}, \bibinfo {author} {\bibfnamefont {Y.}~\bibnamefont {Dusch}},
  \bibinfo {author} {\bibfnamefont {S.}~\bibnamefont {Giordano}}, \bibinfo
  {author} {\bibfnamefont {A.}~\bibnamefont {Klimov}}, \bibinfo {author}
  {\bibfnamefont {V.}~\bibnamefont {Preobrazhensky}}, \ and\ \bibinfo {author}
  {\bibfnamefont {P.}~\bibnamefont {Pernod}},\ }\bibfield  {title} {\enquote
  {\bibinfo {title} {Strain mediated magnetoelectric memory},}\ }\href
  {\doibase 10.1002/9781118869239.ch8} {\bibfield  {journal} {\bibinfo
  {journal} {Nanomagnetic and Spintronic Devices for Energy-Efficient Memory
  and Computing}\ ,\ \bibinfo {pages} {221}} (\bibinfo {year}
  {2016})}\BibitemShut {NoStop}%
\bibitem [{\citenamefont {Yi}\ \emph {et~al.}(2015{\natexlab{b}})\citenamefont
  {Yi}, \citenamefont {Xu},\ and\ \citenamefont {Gross}}]{5yi2015mechanically}%
  \BibitemOpen
  \bibfield  {author} {\bibinfo {author} {\bibfnamefont {M.}~\bibnamefont
  {Yi}}, \bibinfo {author} {\bibfnamefont {B.-X.}\ \bibnamefont {Xu}}, \ and\
  \bibinfo {author} {\bibfnamefont {D.}~\bibnamefont {Gross}},\ }\bibfield
  {title} {\enquote {\bibinfo {title} {Mechanically induced deterministic
  180$^\circ$ switching in nanomagnets},}\ }\href {\doibase
  10.1016/j.mechmat.2015.04.006} {\bibfield  {journal} {\bibinfo  {journal}
  {Mech. Mater.}\ }\textbf {\bibinfo {volume} {87}},\ \bibinfo {pages} {40--49}
  (\bibinfo {year} {2015}{\natexlab{b}})}\BibitemShut {NoStop}%
\bibitem [{\citenamefont {Hu}\ \emph {et~al.}(2016)\citenamefont {Hu},
  \citenamefont {Chen},\ and\ \citenamefont {Nan}}]{6hu2016multiferroic}%
  \BibitemOpen
  \bibfield  {author} {\bibinfo {author} {\bibfnamefont {J.-M.}\ \bibnamefont
  {Hu}}, \bibinfo {author} {\bibfnamefont {L.-Q.}\ \bibnamefont {Chen}}, \ and\
  \bibinfo {author} {\bibfnamefont {C.-W.}\ \bibnamefont {Nan}},\ }\bibfield
  {title} {\enquote {\bibinfo {title} {Multiferroic heterostructures
  integrating ferroelectric and magnetic materials},}\ }\href {\doibase
  10.1002/adma.201502824} {\bibfield  {journal} {\bibinfo  {journal} {Adv.
  Mater.}\ }\textbf {\bibinfo {volume} {28}},\ \bibinfo {pages} {15--39}
  (\bibinfo {year} {2016})}\BibitemShut {NoStop}%
\bibitem [{\citenamefont {Yi}\ \emph {et~al.}(2015{\natexlab{c}})\citenamefont
  {Yi}, \citenamefont {Xu},\ and\ \citenamefont {Shen}}]{7yi2015180}%
  \BibitemOpen
  \bibfield  {author} {\bibinfo {author} {\bibfnamefont {M.}~\bibnamefont
  {Yi}}, \bibinfo {author} {\bibfnamefont {B.-X.}\ \bibnamefont {Xu}}, \ and\
  \bibinfo {author} {\bibfnamefont {Z.}~\bibnamefont {Shen}},\ }\bibfield
  {title} {\enquote {\bibinfo {title} {180$^\circ$ magnetization switching in
  nanocylinders by a mechanical strain},}\ }\href {\doibase
  10.1016/j.eml.2015.03.004} {\bibfield  {journal} {\bibinfo  {journal}
  {Extreme Mech. Lett.}\ }\textbf {\bibinfo {volume} {3}},\ \bibinfo {pages}
  {66--71} (\bibinfo {year} {2015}{\natexlab{c}})}\BibitemShut {NoStop}%
\bibitem [{\citenamefont {Woodcock}\ \emph {et~al.}(2012)\citenamefont
  {Woodcock}, \citenamefont {Zhang}, \citenamefont {Hrkac}, \citenamefont
  {Ciuta}, \citenamefont {Dempsey}, \citenamefont {Schrefl}, \citenamefont
  {Gutfleisch},\ and\ \citenamefont {Givord}}]{8woodcock2012understanding}%
  \BibitemOpen
  \bibfield  {author} {\bibinfo {author} {\bibfnamefont {T.~G.}\ \bibnamefont
  {Woodcock}}, \bibinfo {author} {\bibfnamefont {Y.}~\bibnamefont {Zhang}},
  \bibinfo {author} {\bibfnamefont {G.}~\bibnamefont {Hrkac}}, \bibinfo
  {author} {\bibfnamefont {G.}~\bibnamefont {Ciuta}}, \bibinfo {author}
  {\bibfnamefont {N.~M.}\ \bibnamefont {Dempsey}}, \bibinfo {author}
  {\bibfnamefont {T.}~\bibnamefont {Schrefl}}, \bibinfo {author} {\bibfnamefont
  {O.}~\bibnamefont {Gutfleisch}}, \ and\ \bibinfo {author} {\bibfnamefont
  {D.}~\bibnamefont {Givord}},\ }\bibfield  {title} {\enquote {\bibinfo {title}
  {Understanding the microstructure and coercivity of high performance
  {NdFeB}-based magnets},}\ }\href {\doibase 10.1016/j.scriptamat.2012.05.038}
  {\bibfield  {journal} {\bibinfo  {journal} {Scripta Mater.}\ }\textbf
  {\bibinfo {volume} {67}},\ \bibinfo {pages} {536--541} (\bibinfo {year}
  {2012})}\BibitemShut {NoStop}%
\bibitem [{\citenamefont {Hrkac}\ \emph
  {et~al.}(2014{\natexlab{a}})\citenamefont {Hrkac}, \citenamefont {Woodcock},
  \citenamefont {Butler}, \citenamefont {Saharan}, \citenamefont {Bryan},
  \citenamefont {Schrefl},\ and\ \citenamefont
  {Gutfleisch}}]{9hrkac2014impact}%
  \BibitemOpen
  \bibfield  {author} {\bibinfo {author} {\bibfnamefont {G.}~\bibnamefont
  {Hrkac}}, \bibinfo {author} {\bibfnamefont {T.~G.}\ \bibnamefont {Woodcock}},
  \bibinfo {author} {\bibfnamefont {K.~T.}\ \bibnamefont {Butler}}, \bibinfo
  {author} {\bibfnamefont {L.}~\bibnamefont {Saharan}}, \bibinfo {author}
  {\bibfnamefont {M.~T.}\ \bibnamefont {Bryan}}, \bibinfo {author}
  {\bibfnamefont {T.}~\bibnamefont {Schrefl}}, \ and\ \bibinfo {author}
  {\bibfnamefont {O.}~\bibnamefont {Gutfleisch}},\ }\bibfield  {title}
  {\enquote {\bibinfo {title} {Impact of different {Nd}-rich crystal-phases on
  the coercivity of {Nd--Fe--B} grain ensembles},}\ }\href {\doibase
  10.1016/j.scriptamat.2013.08.029} {\bibfield  {journal} {\bibinfo  {journal}
  {Scripta Mater.}\ }\textbf {\bibinfo {volume} {70}},\ \bibinfo {pages}
  {35--38} (\bibinfo {year} {2014}{\natexlab{a}})}\BibitemShut {NoStop}%
\bibitem [{\citenamefont {Kubo}\ \emph {et~al.}(2014)\citenamefont {Kubo},
  \citenamefont {Wang},\ and\ \citenamefont {Umeno}}]{10kubo2014development}%
  \BibitemOpen
  \bibfield  {author} {\bibinfo {author} {\bibfnamefont {A.}~\bibnamefont
  {Kubo}}, \bibinfo {author} {\bibfnamefont {J.}~\bibnamefont {Wang}}, \ and\
  \bibinfo {author} {\bibfnamefont {Y.}~\bibnamefont {Umeno}},\ }\bibfield
  {title} {\enquote {\bibinfo {title} {Development of interatomic potential for
  {Nd}--{Fe}--{B} permanent magnet and evaluation of magnetic anisotropy near
  the interface and grain boundary},}\ }\href {\doibase
  10.1088/0965-0393/22/6/065014} {\bibfield  {journal} {\bibinfo  {journal}
  {Modell. Simul. Mater. Sci. Eng.}\ }\textbf {\bibinfo {volume} {22}},\
  \bibinfo {pages} {065014} (\bibinfo {year} {2014})}\BibitemShut {NoStop}%
\bibitem [{\citenamefont {Hrkac}\ \emph
  {et~al.}(2014{\natexlab{b}})\citenamefont {Hrkac}, \citenamefont {Butler},
  \citenamefont {Woodcock}, \citenamefont {Saharan}, \citenamefont {Schrefl},\
  and\ \citenamefont {Gutfleisch}}]{11hrkac2014modeling}%
  \BibitemOpen
  \bibfield  {author} {\bibinfo {author} {\bibfnamefont {G.}~\bibnamefont
  {Hrkac}}, \bibinfo {author} {\bibfnamefont {K.}~\bibnamefont {Butler}},
  \bibinfo {author} {\bibfnamefont {T.~G.}\ \bibnamefont {Woodcock}}, \bibinfo
  {author} {\bibfnamefont {L.}~\bibnamefont {Saharan}}, \bibinfo {author}
  {\bibfnamefont {T.}~\bibnamefont {Schrefl}}, \ and\ \bibinfo {author}
  {\bibfnamefont {O.}~\bibnamefont {Gutfleisch}},\ }\bibfield  {title}
  {\enquote {\bibinfo {title} {Modeling of {Nd}-oxide grain boundary phases in
  {Nd}--{Fe}--{B} sintered magnets},}\ }\href {\doibase
  10.1007/s11837-014-0980-5} {\bibfield  {journal} {\bibinfo  {journal} {JOM}\
  }\textbf {\bibinfo {volume} {66}},\ \bibinfo {pages} {1138--1143} (\bibinfo
  {year} {2014}{\natexlab{b}})}\BibitemShut {NoStop}%
\bibitem [{\citenamefont {Hrkac}\ \emph {et~al.}(2010)\citenamefont {Hrkac},
  \citenamefont {Woodcock}, \citenamefont {Freeman}, \citenamefont {Goncharov},
  \citenamefont {Dean}, \citenamefont {Schrefl},\ and\ \citenamefont
  {Gutfleisch}}]{12hrkac2010role}%
  \BibitemOpen
  \bibfield  {author} {\bibinfo {author} {\bibfnamefont {G.}~\bibnamefont
  {Hrkac}}, \bibinfo {author} {\bibfnamefont {T.~G.}\ \bibnamefont {Woodcock}},
  \bibinfo {author} {\bibfnamefont {C.}~\bibnamefont {Freeman}}, \bibinfo
  {author} {\bibfnamefont {A.}~\bibnamefont {Goncharov}}, \bibinfo {author}
  {\bibfnamefont {J.}~\bibnamefont {Dean}}, \bibinfo {author} {\bibfnamefont
  {T.}~\bibnamefont {Schrefl}}, \ and\ \bibinfo {author} {\bibfnamefont
  {O.}~\bibnamefont {Gutfleisch}},\ }\bibfield  {title} {\enquote {\bibinfo
  {title} {The role of local anisotropy profiles at grain boundaries on the
  coercivity of {Nd$_2$Fe$_{14}$B} magnets},}\ }\href {\doibase
  10.1063/1.3519906} {\bibfield  {journal} {\bibinfo  {journal} {Appl. Phys.
  Lett.}\ }\textbf {\bibinfo {volume} {97}},\ \bibinfo {pages} {232511}
  (\bibinfo {year} {2010})}\BibitemShut {NoStop}%
\bibitem [{\citenamefont {Bance}\ \emph
  {et~al.}(2014{\natexlab{a}})\citenamefont {Bance}, \citenamefont {Oezelt},
  \citenamefont {Schrefl}, \citenamefont {Ciuta}, \citenamefont {Dempsey},
  \citenamefont {Givord}, \citenamefont {Winklhofer}, \citenamefont {Hrkac},
  \citenamefont {Zimanyi}, \citenamefont {Gutfleisch}, \citenamefont
  {Woodcock}, \citenamefont {Shoji}, \citenamefont {Yano}, \citenamefont
  {Kato},\ and\ \citenamefont {Manabe}}]{13bance2014influence}%
  \BibitemOpen
  \bibfield  {author} {\bibinfo {author} {\bibfnamefont {S.}~\bibnamefont
  {Bance}}, \bibinfo {author} {\bibfnamefont {H.}~\bibnamefont {Oezelt}},
  \bibinfo {author} {\bibfnamefont {T.}~\bibnamefont {Schrefl}}, \bibinfo
  {author} {\bibfnamefont {G.}~\bibnamefont {Ciuta}}, \bibinfo {author}
  {\bibfnamefont {N.~M.}\ \bibnamefont {Dempsey}}, \bibinfo {author}
  {\bibfnamefont {D.}~\bibnamefont {Givord}}, \bibinfo {author} {\bibfnamefont
  {M.}~\bibnamefont {Winklhofer}}, \bibinfo {author} {\bibfnamefont
  {G.}~\bibnamefont {Hrkac}}, \bibinfo {author} {\bibfnamefont
  {G.}~\bibnamefont {Zimanyi}}, \bibinfo {author} {\bibfnamefont
  {O.}~\bibnamefont {Gutfleisch}}, \bibinfo {author} {\bibfnamefont {T.~G.}\
  \bibnamefont {Woodcock}}, \bibinfo {author} {\bibfnamefont {T.}~\bibnamefont
  {Shoji}}, \bibinfo {author} {\bibfnamefont {M.}~\bibnamefont {Yano}},
  \bibinfo {author} {\bibfnamefont {A}~\bibnamefont {Kato}}, \ and\ \bibinfo
  {author} {\bibfnamefont {A}~\bibnamefont {Manabe}},\ }\bibfield  {title}
  {\enquote {\bibinfo {title} {Influence of defect thickness on the angular
  dependence of coercivity in rare-earth permanent magnets},}\ }\href {\doibase
  10.1063/1.4876451} {\bibfield  {journal} {\bibinfo  {journal} {Appl. Phys.
  Lett.}\ }\textbf {\bibinfo {volume} {104}},\ \bibinfo {pages} {182408}
  (\bibinfo {year} {2014}{\natexlab{a}})}\BibitemShut {NoStop}%
\bibitem [{\citenamefont {Woodcock}\ \emph {et~al.}(2014)\citenamefont
  {Woodcock}, \citenamefont {Ramasse}, \citenamefont {Hrkac}, \citenamefont
  {Shoji}, \citenamefont {Yano}, \citenamefont {Kato},\ and\ \citenamefont
  {Gutfleisch}}]{14woodcock2014atomic}%
  \BibitemOpen
  \bibfield  {author} {\bibinfo {author} {\bibfnamefont {T.~G.}\ \bibnamefont
  {Woodcock}}, \bibinfo {author} {\bibfnamefont {Q.~M.}\ \bibnamefont
  {Ramasse}}, \bibinfo {author} {\bibfnamefont {G.}~\bibnamefont {Hrkac}},
  \bibinfo {author} {\bibfnamefont {T.}~\bibnamefont {Shoji}}, \bibinfo
  {author} {\bibfnamefont {M.}~\bibnamefont {Yano}}, \bibinfo {author}
  {\bibfnamefont {A.}~\bibnamefont {Kato}}, \ and\ \bibinfo {author}
  {\bibfnamefont {O.}~\bibnamefont {Gutfleisch}},\ }\bibfield  {title}
  {\enquote {\bibinfo {title} {Atomic-scale features of phase boundaries in hot
  deformed {Nd}--{Fe}--{Co}--{B}--{Ga} magnets infiltrated with a {Nd}--{Cu}
  eutectic liquid},}\ }\href {\doibase 10.1016/j.actamat.2014.05.045}
  {\bibfield  {journal} {\bibinfo  {journal} {Acta Mater.}\ }\textbf {\bibinfo
  {volume} {77}},\ \bibinfo {pages} {111--124} (\bibinfo {year}
  {2014})}\BibitemShut {NoStop}%
\bibitem [{\citenamefont {Davies}\ \emph {et~al.}(2001)\citenamefont {Davies},
  \citenamefont {Mottram},\ and\ \citenamefont {Harris}}]{15davies2001recent}%
  \BibitemOpen
  \bibfield  {author} {\bibinfo {author} {\bibfnamefont {B.~E.}\ \bibnamefont
  {Davies}}, \bibinfo {author} {\bibfnamefont {R.~S.}\ \bibnamefont {Mottram}},
  \ and\ \bibinfo {author} {\bibfnamefont {I.~R.}\ \bibnamefont {Harris}},\
  }\bibfield  {title} {\enquote {\bibinfo {title} {Recent developments in the
  sintering of {NdFeB}},}\ }\href {\doibase 10.1016/S0254-0584(00)00450-8}
  {\bibfield  {journal} {\bibinfo  {journal} {Mater. Chem. Phys.}\ }\textbf
  {\bibinfo {volume} {67}},\ \bibinfo {pages} {272--281} (\bibinfo {year}
  {2001})}\BibitemShut {NoStop}%
\bibitem [{\citenamefont {Gutfleisch}\ \emph {et~al.}(2011)\citenamefont
  {Gutfleisch}, \citenamefont {Willard}, \citenamefont {Br{\"u}ck},
  \citenamefont {Chen}, \citenamefont {Sankar},\ and\ \citenamefont
  {Liu}}]{16gutfleisch2011magnetic}%
  \BibitemOpen
  \bibfield  {author} {\bibinfo {author} {\bibfnamefont {O.}~\bibnamefont
  {Gutfleisch}}, \bibinfo {author} {\bibfnamefont {M.~A.}\ \bibnamefont
  {Willard}}, \bibinfo {author} {\bibfnamefont {E.}~\bibnamefont {Br{\"u}ck}},
  \bibinfo {author} {\bibfnamefont {C.~H.}\ \bibnamefont {Chen}}, \bibinfo
  {author} {\bibfnamefont {S.~G.}\ \bibnamefont {Sankar}}, \ and\ \bibinfo
  {author} {\bibfnamefont {J.~P.}\ \bibnamefont {Liu}},\ }\bibfield  {title}
  {\enquote {\bibinfo {title} {Magnetic materials and devices for the 21st
  century: stronger, lighter, and more energy efficient},}\ }\href {\doibase
  10.1002/adma.201002180} {\bibfield  {journal} {\bibinfo  {journal} {Adv.
  Mater.}\ }\textbf {\bibinfo {volume} {23}},\ \bibinfo {pages} {821--842}
  (\bibinfo {year} {2011})}\BibitemShut {NoStop}%
\bibitem [{\citenamefont {Sawatzki}\ \emph {et~al.}(2016)\citenamefont
  {Sawatzki}, \citenamefont {K{\"u}bel}, \citenamefont {Ener},\ and\
  \citenamefont {Gutfleisch}}]{new7sawatzki2016grain}%
  \BibitemOpen
  \bibfield  {author} {\bibinfo {author} {\bibfnamefont {S.}~\bibnamefont
  {Sawatzki}}, \bibinfo {author} {\bibfnamefont {C.}~\bibnamefont {K{\"u}bel}},
  \bibinfo {author} {\bibfnamefont {S.}~\bibnamefont {Ener}}, \ and\ \bibinfo
  {author} {\bibfnamefont {O.}~\bibnamefont {Gutfleisch}},\ }\bibfield  {title}
  {\enquote {\bibinfo {title} {Grain boundary diffusion in nanocrystalline
  {Nd--Fe--B} permanent magnets with low-melting eutectics},}\ }\href {\doibase
  10.1016/j.actamat.2016.05.048} {\bibfield  {journal} {\bibinfo  {journal}
  {Acta Mater.}\ }\textbf {\bibinfo {volume} {115}},\ \bibinfo {pages}
  {354--363} (\bibinfo {year} {2016})}\BibitemShut {NoStop}%
\bibitem [{\citenamefont {L{\"o}ewe}\ \emph {et~al.}(2015)\citenamefont
  {L{\"o}ewe}, \citenamefont {Brombacher}, \citenamefont {Katter},\ and\
  \citenamefont {Gutfleisch}}]{new8loewe2015temperature}%
  \BibitemOpen
  \bibfield  {author} {\bibinfo {author} {\bibfnamefont {K.}~\bibnamefont
  {L{\"o}ewe}}, \bibinfo {author} {\bibfnamefont {C.}~\bibnamefont
  {Brombacher}}, \bibinfo {author} {\bibfnamefont {M.}~\bibnamefont {Katter}},
  \ and\ \bibinfo {author} {\bibfnamefont {O.}~\bibnamefont {Gutfleisch}},\
  }\bibfield  {title} {\enquote {\bibinfo {title} {Temperature-dependent {Dy}
  diffusion processes in {Nd--Fe--B} permanent magnets},}\ }\href {\doibase
  10.1016/j.actamat.2014.09.039} {\bibfield  {journal} {\bibinfo  {journal}
  {Acta Mater.}\ }\textbf {\bibinfo {volume} {83}},\ \bibinfo {pages}
  {248--255} (\bibinfo {year} {2015})}\BibitemShut {NoStop}%
\bibitem [{\citenamefont {De~Groot}\ and\ \citenamefont
  {de~Kort}(1999)}]{17de1999magnetoelastic}%
  \BibitemOpen
  \bibfield  {author} {\bibinfo {author} {\bibfnamefont {C.~H.}\ \bibnamefont
  {De~Groot}}\ and\ \bibinfo {author} {\bibfnamefont {K.}~\bibnamefont
  {de~Kort}},\ }\bibfield  {title} {\enquote {\bibinfo {title} {Magnetoelastic
  anisotropy in {NdFeB} permanent magnets},}\ }\href {\doibase
  10.1063/1.370675} {\bibfield  {journal} {\bibinfo  {journal} {J. Appl.
  Phys.}\ }\textbf {\bibinfo {volume} {85}},\ \bibinfo {pages} {8312--8316}
  (\bibinfo {year} {1999})}\BibitemShut {NoStop}%
\bibitem [{\citenamefont {Murakami}\ \emph {et~al.}(2015)\citenamefont
  {Murakami}, \citenamefont {Sasaki}, \citenamefont {Ohkubo},\ and\
  \citenamefont {Hono}}]{18murakami2015strain}%
  \BibitemOpen
  \bibfield  {author} {\bibinfo {author} {\bibfnamefont {Y.}~\bibnamefont
  {Murakami}}, \bibinfo {author} {\bibfnamefont {T.~T.}\ \bibnamefont
  {Sasaki}}, \bibinfo {author} {\bibfnamefont {T.}~\bibnamefont {Ohkubo}}, \
  and\ \bibinfo {author} {\bibfnamefont {K.}~\bibnamefont {Hono}},\ }\bibfield
  {title} {\enquote {\bibinfo {title} {Strain measurements from
  {Nd$_2$Fe$_{14}$B} grains in sintered magnets using artificial moir{\'e}
  fringes},}\ }\href {\doibase 10.1016/j.actamat.2015.08.058} {\bibfield
  {journal} {\bibinfo  {journal} {Acta Mater.}\ }\textbf {\bibinfo {volume}
  {101}},\ \bibinfo {pages} {101--106} (\bibinfo {year} {2015})}\BibitemShut
  {NoStop}%
\bibitem [{\citenamefont {Murakami}\ \emph {et~al.}(2016)\citenamefont
  {Murakami}, \citenamefont {Niitsu}, \citenamefont {Kaneko}, \citenamefont
  {Tanigaki}, \citenamefont {Sasaki}, \citenamefont {Akase}, \citenamefont
  {Shindo}, \citenamefont {Ohkubo},\ and\ \citenamefont
  {Hono}}]{new10murakami2016strain}%
  \BibitemOpen
  \bibfield  {author} {\bibinfo {author} {\bibfnamefont {Y.}~\bibnamefont
  {Murakami}}, \bibinfo {author} {\bibfnamefont {K.}~\bibnamefont {Niitsu}},
  \bibinfo {author} {\bibfnamefont {S.}~\bibnamefont {Kaneko}}, \bibinfo
  {author} {\bibfnamefont {T.}~\bibnamefont {Tanigaki}}, \bibinfo {author}
  {\bibfnamefont {T.}~\bibnamefont {Sasaki}}, \bibinfo {author} {\bibfnamefont
  {Z.}~\bibnamefont {Akase}}, \bibinfo {author} {\bibfnamefont
  {D.}~\bibnamefont {Shindo}}, \bibinfo {author} {\bibfnamefont
  {T.}~\bibnamefont {Ohkubo}}, \ and\ \bibinfo {author} {\bibfnamefont
  {K.}~\bibnamefont {Hono}},\ }\bibfield  {title} {\enquote {\bibinfo {title}
  {Strain measurement in ferromagnetic crystals using dark-field electron
  holography},}\ }\href {\doibase 10.1063/1.4967005} {\bibfield  {journal}
  {\bibinfo  {journal} {Appl. Phys. Lett.}\ }\textbf {\bibinfo {volume}
  {109}},\ \bibinfo {pages} {193102} (\bibinfo {year} {2016})}\BibitemShut
  {NoStop}%
\bibitem [{\citenamefont {Moriya}\ \emph {et~al.}(2009)\citenamefont {Moriya},
  \citenamefont {Tsuchiura},\ and\ \citenamefont {Sakuma}}]{19moriya2009first}%
  \BibitemOpen
  \bibfield  {author} {\bibinfo {author} {\bibfnamefont {H.}~\bibnamefont
  {Moriya}}, \bibinfo {author} {\bibfnamefont {H.}~\bibnamefont {Tsuchiura}}, \
  and\ \bibinfo {author} {\bibfnamefont {A.}~\bibnamefont {Sakuma}},\
  }\bibfield  {title} {\enquote {\bibinfo {title} {First principles calculation
  of crystal field parameter near surfaces of {Nd$_2$Fe$_{14}$B}},}\ }\href
  {\doibase 10.1063/1.3073931} {\bibfield  {journal} {\bibinfo  {journal} {J.
  Appl. Phys.}\ }\textbf {\bibinfo {volume} {105}},\ \bibinfo {pages} {07A740}
  (\bibinfo {year} {2009})}\BibitemShut {NoStop}%
\bibitem [{\citenamefont {Suzuki}\ \emph {et~al.}(2014)\citenamefont {Suzuki},
  \citenamefont {Toga},\ and\ \citenamefont {Sakuma}}]{20suzuki2014effects}%
  \BibitemOpen
  \bibfield  {author} {\bibinfo {author} {\bibfnamefont {T.}~\bibnamefont
  {Suzuki}}, \bibinfo {author} {\bibfnamefont {Y.}~\bibnamefont {Toga}}, \ and\
  \bibinfo {author} {\bibfnamefont {A.}~\bibnamefont {Sakuma}},\ }\bibfield
  {title} {\enquote {\bibinfo {title} {Effects of deformation on the crystal
  field parameter of the {Nd} ions in {Nd$_2$Fe$_{14}$B}},}\ }\href {\doibase
  10.1063/1.4860937} {\bibfield  {journal} {\bibinfo  {journal} {J. Appl.
  Phys.}\ }\textbf {\bibinfo {volume} {115}},\ \bibinfo {pages} {17A703}
  (\bibinfo {year} {2014})}\BibitemShut {NoStop}%
\bibitem [{\citenamefont {Tanaka}\ \emph
  {et~al.}(2011{\natexlab{a}})\citenamefont {Tanaka}, \citenamefont {Moriya},
  \citenamefont {Tsuchiura}, \citenamefont {Sakuma}, \citenamefont
  {Divi{\v{s}}},\ and\ \citenamefont {Nov{\'a}k}}]{21tanaka2011first}%
  \BibitemOpen
  \bibfield  {author} {\bibinfo {author} {\bibfnamefont {S.}~\bibnamefont
  {Tanaka}}, \bibinfo {author} {\bibfnamefont {H.}~\bibnamefont {Moriya}},
  \bibinfo {author} {\bibfnamefont {H.}~\bibnamefont {Tsuchiura}}, \bibinfo
  {author} {\bibfnamefont {A.}~\bibnamefont {Sakuma}}, \bibinfo {author}
  {\bibfnamefont {M.}~\bibnamefont {Divi{\v{s}}}}, \ and\ \bibinfo {author}
  {\bibfnamefont {P.}~\bibnamefont {Nov{\'a}k}},\ }\bibfield  {title} {\enquote
  {\bibinfo {title} {First principles study on the local magnetic anisotropy
  near surfaces of {Dy$_2$Fe$_{14}$B} and {Nd$_2$Fe$_{14}$B} magnets},}\ }\href
  {\doibase 10.1063/1.3553935} {\bibfield  {journal} {\bibinfo  {journal} {J.
  Appl. Phys.}\ }\textbf {\bibinfo {volume} {109}},\ \bibinfo {pages} {07A702}
  (\bibinfo {year} {2011}{\natexlab{a}})}\BibitemShut {NoStop}%
\bibitem [{\citenamefont {Toga}\ \emph {et~al.}(2015)\citenamefont {Toga},
  \citenamefont {Suzuki},\ and\ \citenamefont {Sakuma}}]{22tanaka2011first}%
  \BibitemOpen
  \bibfield  {author} {\bibinfo {author} {\bibfnamefont {Y.}~\bibnamefont
  {Toga}}, \bibinfo {author} {\bibfnamefont {T.}~\bibnamefont {Suzuki}}, \ and\
  \bibinfo {author} {\bibfnamefont {A.}~\bibnamefont {Sakuma}},\ }\bibfield
  {title} {\enquote {\bibinfo {title} {Effects of trace elements on the crystal
  field parameters of {Nd} ions at the surface of {Nd$_2$Fe$_{14}$B} grains},}\
  }\href {\doibase 10.1063/1.4922500} {\bibfield  {journal} {\bibinfo
  {journal} {J. Appl. Phys.}\ }\textbf {\bibinfo {volume} {117}},\ \bibinfo
  {pages} {223905} (\bibinfo {year} {2015})}\BibitemShut {NoStop}%
\bibitem [{\citenamefont {Tanaka}\ \emph
  {et~al.}(2011{\natexlab{b}})\citenamefont {Tanaka}, \citenamefont {Moriya},
  \citenamefont {Tsuchiura}, \citenamefont {Sakuma}, \citenamefont
  {Divi{\v{s}}},\ and\ \citenamefont {Nov{\'a}k}}]{23tanaka2011first}%
  \BibitemOpen
  \bibfield  {author} {\bibinfo {author} {\bibfnamefont {S.}~\bibnamefont
  {Tanaka}}, \bibinfo {author} {\bibfnamefont {H.}~\bibnamefont {Moriya}},
  \bibinfo {author} {\bibfnamefont {H.}~\bibnamefont {Tsuchiura}}, \bibinfo
  {author} {\bibfnamefont {A.}~\bibnamefont {Sakuma}}, \bibinfo {author}
  {\bibfnamefont {M.}~\bibnamefont {Divi{\v{s}}}}, \ and\ \bibinfo {author}
  {\bibfnamefont {P.}~\bibnamefont {Nov{\'a}k}},\ }\bibfield  {title} {\enquote
  {\bibinfo {title} {First-principles calculation of crystal field parameters
  of {Dy} ions substituted for {Nd} in {Nd}--{Fe}--{B} magnets},}\ }\href
  {\doibase 10.1088/1742-6596/266/1/012045} {\bibfield  {journal} {\bibinfo
  {journal} {J. Phys. Conf. Ser.}\ }\textbf {\bibinfo {volume} {266}},\
  \bibinfo {pages} {012045} (\bibinfo {year} {2011}{\natexlab{b}})}\BibitemShut
  {NoStop}%
\bibitem [{\citenamefont {Tatetsu}\ \emph {et~al.}(2016)\citenamefont
  {Tatetsu}, \citenamefont {Tsuneyuki},\ and\ \citenamefont
  {Gohda}}]{new11tatetsu2016first}%
  \BibitemOpen
  \bibfield  {author} {\bibinfo {author} {\bibfnamefont {Y.}~\bibnamefont
  {Tatetsu}}, \bibinfo {author} {\bibfnamefont {S.}~\bibnamefont {Tsuneyuki}},
  \ and\ \bibinfo {author} {\bibfnamefont {Y.}~\bibnamefont {Gohda}},\
  }\bibfield  {title} {\enquote {\bibinfo {title} {First-principles study of
  the role of {Cu} in improving the coercivity of {Nd-Fe-B} permanent
  magnets},}\ }\href {\doibase 10.1103/PhysRevApplied.6.064029} {\bibfield
  {journal} {\bibinfo  {journal} {Phys. Rev. Appl.}\ }\textbf {\bibinfo
  {volume} {6}},\ \bibinfo {pages} {064029} (\bibinfo {year}
  {2016})}\BibitemShut {NoStop}%
\bibitem [{\citenamefont {Drebov}\ \emph {et~al.}(2013)\citenamefont {Drebov},
  \citenamefont {Martinez-Limia}, \citenamefont {Kunz}, \citenamefont {Gola},
  \citenamefont {Shigematsu}, \citenamefont {Eckl}, \citenamefont {Gumbsch},\
  and\ \citenamefont {Els{\"a}sser}}]{24drebov2013ab}%
  \BibitemOpen
  \bibfield  {author} {\bibinfo {author} {\bibfnamefont {N.}~\bibnamefont
  {Drebov}}, \bibinfo {author} {\bibfnamefont {A.}~\bibnamefont
  {Martinez-Limia}}, \bibinfo {author} {\bibfnamefont {L.}~\bibnamefont
  {Kunz}}, \bibinfo {author} {\bibfnamefont {A.}~\bibnamefont {Gola}}, \bibinfo
  {author} {\bibfnamefont {T.}~\bibnamefont {Shigematsu}}, \bibinfo {author}
  {\bibfnamefont {T.}~\bibnamefont {Eckl}}, \bibinfo {author} {\bibfnamefont
  {P.}~\bibnamefont {Gumbsch}}, \ and\ \bibinfo {author} {\bibfnamefont
  {C.}~\bibnamefont {Els{\"a}sser}},\ }\bibfield  {title} {\enquote {\bibinfo
  {title} {\textit{Ab initio} screening methodology applied to the search for
  new permanent magnetic materials},}\ }\href {\doibase
  10.1088/1367-2630/15/12/125023} {\bibfield  {journal} {\bibinfo  {journal}
  {New J. Phys.}\ }\textbf {\bibinfo {volume} {15}},\ \bibinfo {pages} {125023}
  (\bibinfo {year} {2013})}\BibitemShut {NoStop}%
\bibitem [{\citenamefont {Kitagawa}\ and\ \citenamefont
  {Asari}(2010)}]{25kitagawa2010magnetic}%
  \BibitemOpen
  \bibfield  {author} {\bibinfo {author} {\bibfnamefont {I.}~\bibnamefont
  {Kitagawa}}\ and\ \bibinfo {author} {\bibfnamefont {Y.}~\bibnamefont
  {Asari}},\ }\bibfield  {title} {\enquote {\bibinfo {title} {Magnetic
  anisotropy of {R$_2$Fe$_{14}$B} ({R=Nd, Gd, Y}): Density functional
  calculation by using the linear combination of pseudo-atomic-orbital
  method},}\ }\href {\doibase 10.1103/PhysRevB.81.214408} {\bibfield  {journal}
  {\bibinfo  {journal} {Phys. Rev. B}\ }\textbf {\bibinfo {volume} {81}},\
  \bibinfo {pages} {214408} (\bibinfo {year} {2010})}\BibitemShut {NoStop}%
\bibitem [{\citenamefont {Asali}\ \emph {et~al.}(2014)\citenamefont {Asali},
  \citenamefont {Toson}, \citenamefont {Blaha},\ and\ \citenamefont
  {Fidler}}]{new13asali2014dependence}%
  \BibitemOpen
  \bibfield  {author} {\bibinfo {author} {\bibfnamefont {A.}~\bibnamefont
  {Asali}}, \bibinfo {author} {\bibfnamefont {P.}~\bibnamefont {Toson}},
  \bibinfo {author} {\bibfnamefont {P.}~\bibnamefont {Blaha}}, \ and\ \bibinfo
  {author} {\bibfnamefont {J.}~\bibnamefont {Fidler}},\ }\bibfield  {title}
  {\enquote {\bibinfo {title} {Dependence of magnetic anisotropy energy on c/a
  ratio of {X$_2$Fe$_{14}$B (X= Y,Pr,Dy)}},}\ }\href {\doibase
  10.1109/TMAG.2014.2326431} {\bibfield  {journal} {\bibinfo  {journal} {IEEE
  Trans. Magn.}\ }\textbf {\bibinfo {volume} {50}},\ \bibinfo {pages} {1--4}
  (\bibinfo {year} {2014})}\BibitemShut {NoStop}%
\bibitem [{\citenamefont {Torbatian}\ \emph {et~al.}(2014)\citenamefont
  {Torbatian}, \citenamefont {Ozaki}, \citenamefont {Tsuneyuki},\ and\
  \citenamefont {Gohda}}]{torbatian2014strain}%
  \BibitemOpen
  \bibfield  {author} {\bibinfo {author} {\bibfnamefont {Z.}~\bibnamefont
  {Torbatian}}, \bibinfo {author} {\bibfnamefont {T.}~\bibnamefont {Ozaki}},
  \bibinfo {author} {\bibfnamefont {S.}~\bibnamefont {Tsuneyuki}}, \ and\
  \bibinfo {author} {\bibfnamefont {Y.}~\bibnamefont {Gohda}},\ }\bibfield
  {title} {\enquote {\bibinfo {title} {Strain effects on the magnetic
  anisotropy of {Y$_2$Fe$_{14}$B} examined by first-principles calculations},}\
  }\href {\doibase 10.1063/1.4883840} {\bibfield  {journal} {\bibinfo
  {journal} {Appl. Phys. Lett.}\ }\textbf {\bibinfo {volume} {104}},\ \bibinfo
  {pages} {242403} (\bibinfo {year} {2014})}\BibitemShut {NoStop}%
\bibitem [{\citenamefont {Kresse}\ and\ \citenamefont
  {Furthm{\"u}ller}(1996)}]{26kresse1996efficient}%
  \BibitemOpen
  \bibfield  {author} {\bibinfo {author} {\bibfnamefont {G.}~\bibnamefont
  {Kresse}}\ and\ \bibinfo {author} {\bibfnamefont {J.}~\bibnamefont
  {Furthm{\"u}ller}},\ }\bibfield  {title} {\enquote {\bibinfo {title}
  {Efficient iterative schemes for \textit{ab initio} total-energy calculations
  using a plane-wave basis set},}\ }\href {\doibase 10.1103/PhysRevB.54.11169}
  {\bibfield  {journal} {\bibinfo  {journal} {Phys. Rev. B}\ }\textbf {\bibinfo
  {volume} {54}},\ \bibinfo {pages} {11169} (\bibinfo {year}
  {1996})}\BibitemShut {NoStop}%
\bibitem [{\citenamefont {Perdew}\ \emph {et~al.}(1996)\citenamefont {Perdew},
  \citenamefont {Burke},\ and\ \citenamefont
  {Ernzerhof}}]{new3perdew1996generalized}%
  \BibitemOpen
  \bibfield  {author} {\bibinfo {author} {\bibfnamefont {J.~P.}\ \bibnamefont
  {Perdew}}, \bibinfo {author} {\bibfnamefont {K.}~\bibnamefont {Burke}}, \
  and\ \bibinfo {author} {\bibfnamefont {M.}~\bibnamefont {Ernzerhof}},\
  }\bibfield  {title} {\enquote {\bibinfo {title} {Generalized gradient
  approximation made simple},}\ }\href {\doibase 10.1103/PhysRevLett.77.3865}
  {\bibfield  {journal} {\bibinfo  {journal} {Phys. Rev. Lett.}\ }\textbf
  {\bibinfo {volume} {77}},\ \bibinfo {pages} {3865} (\bibinfo {year}
  {1996})}\BibitemShut {NoStop}%
\bibitem [{\citenamefont {Donahue}\ and\ \citenamefont
  {Porter}(2016)}]{27oommf}%
  \BibitemOpen
  \bibfield  {author} {\bibinfo {author} {\bibfnamefont {M.~J.}\ \bibnamefont
  {Donahue}}\ and\ \bibinfo {author} {\bibfnamefont {D.~G}\ \bibnamefont
  {Porter}},\ }\bibfield  {title} {\enquote {\bibinfo {title} {{OOMMF Software
  Package}},}\ }\href@noop {} {\bibfield  {journal} {\bibinfo  {journal}
  {http://math.nist.gov/oommf}\ } (\bibinfo {year} {2016})}\BibitemShut
  {NoStop}%
\bibitem [{\citenamefont {Gilbert}(2004)}]{new4gilbert2004phenomenological}%
  \BibitemOpen
  \bibfield  {author} {\bibinfo {author} {\bibfnamefont {T.~L.}\ \bibnamefont
  {Gilbert}},\ }\bibfield  {title} {\enquote {\bibinfo {title} {A
  phenomenological theory of damping in ferromagnetic materials},}\ }\href
  {\doibase 10.1109/TMAG.2004.836740} {\bibfield  {journal} {\bibinfo
  {journal} {IEEE Trans. Magn.}\ }\textbf {\bibinfo {volume} {40}},\ \bibinfo
  {pages} {3443--3449} (\bibinfo {year} {2004})}\BibitemShut {NoStop}%
\bibitem [{\citenamefont {Yi}\ and\ \citenamefont
  {Xu}(2014)}]{new5yi2014constraint}%
  \BibitemOpen
  \bibfield  {author} {\bibinfo {author} {\bibfnamefont {M.}~\bibnamefont
  {Yi}}\ and\ \bibinfo {author} {\bibfnamefont {B.-X.}\ \bibnamefont {Xu}},\
  }\bibfield  {title} {\enquote {\bibinfo {title} {A constraint-free phase
  field model for ferromagnetic domain evolution},}\ }\href {\doibase
  10.1098/rspa.2014.0517} {\bibfield  {journal} {\bibinfo  {journal} {Proc. R.
  Soc. A}\ }\textbf {\bibinfo {volume} {470}},\ \bibinfo {pages} {20140517}
  (\bibinfo {year} {2014})}\BibitemShut {NoStop}%
\bibitem [{\citenamefont {Yi}\ and\ \citenamefont {Xu}(2016)}]{new9yi2016real}%
  \BibitemOpen
  \bibfield  {author} {\bibinfo {author} {\bibfnamefont {M.}~\bibnamefont
  {Yi}}\ and\ \bibinfo {author} {\bibfnamefont {B.-X.}\ \bibnamefont {Xu}},\
  }\bibfield  {title} {\enquote {\bibinfo {title} {A real-space and
  constraint-free phase field model for the microstructure of ferromagnetic
  shape memory alloys},}\ }\href {\doibase 10.1007/s10704-016-0152-4}
  {\bibfield  {journal} {\bibinfo  {journal} {Int. J. Fract.}\ }\textbf
  {\bibinfo {volume} {202}},\ \bibinfo {pages} {179--194} (\bibinfo {year}
  {2016})}\BibitemShut {NoStop}%
\bibitem [{\citenamefont {Sepehri-Amin}\ \emph {et~al.}(2014)\citenamefont
  {Sepehri-Amin}, \citenamefont {Ohkubo}, \citenamefont {Gruber}, \citenamefont
  {Schrefl},\ and\ \citenamefont {Hono}}]{28sepehri2014micromagnetic}%
  \BibitemOpen
  \bibfield  {author} {\bibinfo {author} {\bibfnamefont {H.}~\bibnamefont
  {Sepehri-Amin}}, \bibinfo {author} {\bibfnamefont {T.}~\bibnamefont
  {Ohkubo}}, \bibinfo {author} {\bibfnamefont {M.}~\bibnamefont {Gruber}},
  \bibinfo {author} {\bibfnamefont {T.}~\bibnamefont {Schrefl}}, \ and\
  \bibinfo {author} {\bibfnamefont {K.}~\bibnamefont {Hono}},\ }\bibfield
  {title} {\enquote {\bibinfo {title} {Micromagnetic simulations on the grain
  size dependence of coercivity in anisotropic {Nd}--{Fe}--{B} sintered
  magnets},}\ }\href {\doibase 10.1016/j.scriptamat.2014.06.020} {\bibfield
  {journal} {\bibinfo  {journal} {Scripta Mater.}\ }\textbf {\bibinfo {volume}
  {89}},\ \bibinfo {pages} {29--32} (\bibinfo {year} {2014})}\BibitemShut
  {NoStop}%
\bibitem [{\citenamefont {Koon}\ \emph {et~al.}(1985)\citenamefont {Koon},
  \citenamefont {Das}, \citenamefont {Rubinstein},\ and\ \citenamefont
  {Tyson}}]{29koon1985magnetic}%
  \BibitemOpen
  \bibfield  {author} {\bibinfo {author} {\bibfnamefont {N.~C.}\ \bibnamefont
  {Koon}}, \bibinfo {author} {\bibfnamefont {B.~N.}\ \bibnamefont {Das}},
  \bibinfo {author} {\bibfnamefont {M.}~\bibnamefont {Rubinstein}}, \ and\
  \bibinfo {author} {\bibfnamefont {J.}~\bibnamefont {Tyson}},\ }\bibfield
  {title} {\enquote {\bibinfo {title} {Magnetic properties of {R$_2$Fe$_{14}$B}
  single crystals},}\ }\href {\doibase 10.1063/1.334681} {\bibfield  {journal}
  {\bibinfo  {journal} {J. Appl. Phys.}\ }\textbf {\bibinfo {volume} {57}},\
  \bibinfo {pages} {4091--4093} (\bibinfo {year} {1985})}\BibitemShut {NoStop}%
\bibitem [{\citenamefont {Sagawa}\ \emph {et~al.}(1985)\citenamefont {Sagawa},
  \citenamefont {Fujimura}, \citenamefont {Yamamoto}, \citenamefont
  {Matsuura},\ and\ \citenamefont {Hirosawa}}]{30sagawa1985magnetic}%
  \BibitemOpen
  \bibfield  {author} {\bibinfo {author} {\bibfnamefont {M.}~\bibnamefont
  {Sagawa}}, \bibinfo {author} {\bibfnamefont {S.}~\bibnamefont {Fujimura}},
  \bibinfo {author} {\bibfnamefont {H.}~\bibnamefont {Yamamoto}}, \bibinfo
  {author} {\bibfnamefont {Y.}~\bibnamefont {Matsuura}}, \ and\ \bibinfo
  {author} {\bibfnamefont {S.}~\bibnamefont {Hirosawa}},\ }\bibfield  {title}
  {\enquote {\bibinfo {title} {Magnetic properties of rare-earth-iron-boron
  permanent magnet materials},}\ }\href {\doibase 10.1063/1.334629} {\bibfield
  {journal} {\bibinfo  {journal} {J. Appl. Phys.}\ }\textbf {\bibinfo {volume}
  {57}},\ \bibinfo {pages} {4094--4096} (\bibinfo {year} {1985})}\BibitemShut
  {NoStop}%
\bibitem [{\citenamefont {Yamada}\ \emph {et~al.}(1986)\citenamefont {Yamada},
  \citenamefont {Tokuhara}, \citenamefont {Ono}, \citenamefont {Sagawa},\ and\
  \citenamefont {Matsuura}}]{31yamada1986magnetocrystalline}%
  \BibitemOpen
  \bibfield  {author} {\bibinfo {author} {\bibfnamefont {O.}~\bibnamefont
  {Yamada}}, \bibinfo {author} {\bibfnamefont {H.}~\bibnamefont {Tokuhara}},
  \bibinfo {author} {\bibfnamefont {F.}~\bibnamefont {Ono}}, \bibinfo {author}
  {\bibfnamefont {M.}~\bibnamefont {Sagawa}}, \ and\ \bibinfo {author}
  {\bibfnamefont {Y.}~\bibnamefont {Matsuura}},\ }\bibfield  {title} {\enquote
  {\bibinfo {title} {Magnetocrystalline anisotropy in {Nd$_2$Fe$_{14}$B}
  intermetallic compound},}\ }\href {\doibase 10.1016/0304-8853(86)90718-3}
  {\bibfield  {journal} {\bibinfo  {journal} {J. Magn. Magn. Mater.}\ }\textbf
  {\bibinfo {volume} {54}},\ \bibinfo {pages} {585--586} (\bibinfo {year}
  {1986})}\BibitemShut {NoStop}%
\bibitem [{\citenamefont {Otani}\ \emph {et~al.}(1987)\citenamefont {Otani},
  \citenamefont {Miyajima},\ and\ \citenamefont
  {Chikazumi}}]{32otani1987magnetocrystalline}%
  \BibitemOpen
  \bibfield  {author} {\bibinfo {author} {\bibfnamefont {Y.}~\bibnamefont
  {Otani}}, \bibinfo {author} {\bibfnamefont {H.}~\bibnamefont {Miyajima}}, \
  and\ \bibinfo {author} {\bibfnamefont {S.}~\bibnamefont {Chikazumi}},\
  }\bibfield  {title} {\enquote {\bibinfo {title} {Magnetocrystalline
  anisotropy in {Nd}--{Fe}--{B} magnet},}\ }\href {\doibase 10.1063/1.338745}
  {\bibfield  {journal} {\bibinfo  {journal} {J. Appl. Phys.}\ }\textbf
  {\bibinfo {volume} {61}},\ \bibinfo {pages} {3436--3438} (\bibinfo {year}
  {1987})}\BibitemShut {NoStop}%
\bibitem [{\citenamefont {Givord}\ \emph {et~al.}(1985)\citenamefont {Givord},
  \citenamefont {Li}, \citenamefont {Moreau}, \citenamefont {de~la
  B{\^a}thie},\ and\ \citenamefont {de~Lacheisserie}}]{33givord1985structural}%
  \BibitemOpen
  \bibfield  {author} {\bibinfo {author} {\bibfnamefont {D.}~\bibnamefont
  {Givord}}, \bibinfo {author} {\bibfnamefont {H.~S.}\ \bibnamefont {Li}},
  \bibinfo {author} {\bibfnamefont {J.~M.}\ \bibnamefont {Moreau}}, \bibinfo
  {author} {\bibfnamefont {R.~P.}\ \bibnamefont {de~la B{\^a}thie}}, \ and\
  \bibinfo {author} {\bibfnamefont {E.~D.~T.}\ \bibnamefont
  {de~Lacheisserie}},\ }\bibfield  {title} {\enquote {\bibinfo {title}
  {Structural and magnetic properties in {R$_2$Fe$_{14}$B} compounds},}\ }\href
  {\doibase 10.1016/0378-4363(85)90251-7} {\bibfield  {journal} {\bibinfo
  {journal} {Physica B+ C}\ }\textbf {\bibinfo {volume} {130}},\ \bibinfo
  {pages} {323--326} (\bibinfo {year} {1985})}\BibitemShut {NoStop}%
\bibitem [{\citenamefont {Griffith}(1921)}]{griffith1921phenomena}%
  \BibitemOpen
  \bibfield  {author} {\bibinfo {author} {\bibfnamefont {A.~A.}\ \bibnamefont
  {Griffith}},\ }\bibfield  {title} {\enquote {\bibinfo {title} {The phenomena
  of rupture and flow in solids},}\ }\href {\doibase 10.1098/rsta.1921.0006}
  {\bibfield  {journal} {\bibinfo  {journal} {Philos. Trans. R. Soc. London
  Ser. A}\ }\textbf {\bibinfo {volume} {221}},\ \bibinfo {pages} {163--198}
  (\bibinfo {year} {1921})}\BibitemShut {NoStop}%
\bibitem [{\citenamefont {Herbst}(1991)}]{34herbst1991r}%
  \BibitemOpen
  \bibfield  {author} {\bibinfo {author} {\bibfnamefont {J.~F.}\ \bibnamefont
  {Herbst}},\ }\bibfield  {title} {\enquote {\bibinfo {title}
  {{Nd$_2$Fe$_{14}$B} materials: Intrinsic properties and technological
  aspects},}\ }\href {\doibase 10.1103/RevModPhys.63.819} {\bibfield  {journal}
  {\bibinfo  {journal} {Rev. Mod. Phys.}\ }\textbf {\bibinfo {volume} {63}},\
  \bibinfo {pages} {819} (\bibinfo {year} {1991})}\BibitemShut {NoStop}%
\bibitem [{\citenamefont {Nordstrom}\ \emph {et~al.}(1993)\citenamefont
  {Nordstrom}, \citenamefont {Johansson},\ and\ \citenamefont
  {Brooks}}]{35nordstrom1993calculation}%
  \BibitemOpen
  \bibfield  {author} {\bibinfo {author} {\bibfnamefont {L.}~\bibnamefont
  {Nordstrom}}, \bibinfo {author} {\bibfnamefont {B.}~\bibnamefont
  {Johansson}}, \ and\ \bibinfo {author} {\bibfnamefont {M.~S.~S.}\
  \bibnamefont {Brooks}},\ }\bibfield  {title} {\enquote {\bibinfo {title}
  {Calculation of the electronic structure and the magnetic moments of
  {Nd$_2$Fe$_{14}$B}},}\ }\href {\doibase 10.1088/0953-8984/5/42/008/}
  {\bibfield  {journal} {\bibinfo  {journal} {J. Phys. Condens. Mat.}\ }\textbf
  {\bibinfo {volume} {5}},\ \bibinfo {pages} {7859} (\bibinfo {year}
  {1993})}\BibitemShut {NoStop}%
\bibitem [{\citenamefont {Jaswal}(1990)}]{36jaswal1990electronic}%
  \BibitemOpen
  \bibfield  {author} {\bibinfo {author} {\bibfnamefont {S.~S.}\ \bibnamefont
  {Jaswal}},\ }\bibfield  {title} {\enquote {\bibinfo {title} {Electronic
  structure and magnetism of {R$_2$Fe$_{14}$B ({R=Y, Nd})} compounds},}\ }\href
  {\doibase 10.1103/PhysRevB.41.9697} {\bibfield  {journal} {\bibinfo
  {journal} {Phys. Rev. B}\ }\textbf {\bibinfo {volume} {41}},\ \bibinfo
  {pages} {9697} (\bibinfo {year} {1990})}\BibitemShut {NoStop}%
\bibitem [{\citenamefont {Kitagawa}(2009)}]{37kitagawa2009calculation}%
  \BibitemOpen
  \bibfield  {author} {\bibinfo {author} {\bibfnamefont {I.}~\bibnamefont
  {Kitagawa}},\ }\bibfield  {title} {\enquote {\bibinfo {title} {Calculation of
  electronic structures and magnetic moments of {Nd$_2$Fe$_{14}$B} and
  {Dy$_2$Fe$_{14}$B} by using linear-combination-of-pseudo-atomic-orbital
  method},}\ }\href {\doibase 10.1063/1.3068458} {\bibfield  {journal}
  {\bibinfo  {journal} {J. Appl. Phys.}\ }\textbf {\bibinfo {volume} {105}},\
  \bibinfo {pages} {07E502} (\bibinfo {year} {2009})}\BibitemShut {NoStop}%
\bibitem [{\citenamefont {Gu}\ and\ \citenamefont
  {Ching}(1987)}]{new1gu1987comparative}%
  \BibitemOpen
  \bibfield  {author} {\bibinfo {author} {\bibfnamefont {Z.-Q.}\ \bibnamefont
  {Gu}}\ and\ \bibinfo {author} {\bibfnamefont {W.~Y.}\ \bibnamefont {Ching}},\
  }\bibfield  {title} {\enquote {\bibinfo {title} {Comparative studies of
  electronic and magnetic structures in {Y$_2$Fe$_{14}$B}, {Nd$_2$Fe$_{14}$B},
  {Y$_2$Co$_{14}$B}, and {Nd$_2$Co$_{14}$B}},}\ }\href {\doibase
  10.1103/PhysRevB.36.8530} {\bibfield  {journal} {\bibinfo  {journal} {Phys.
  Rev. B}\ }\textbf {\bibinfo {volume} {36}},\ \bibinfo {pages} {8530}
  (\bibinfo {year} {1987})}\BibitemShut {NoStop}%
\bibitem [{\citenamefont {Ching}\ and\ \citenamefont
  {Gu}(1987)}]{new2ching1987electronic}%
  \BibitemOpen
  \bibfield  {author} {\bibinfo {author} {\bibfnamefont {W.~Y.}\ \bibnamefont
  {Ching}}\ and\ \bibinfo {author} {\bibfnamefont {Z.-Q.}\ \bibnamefont {Gu}},\
  }\bibfield  {title} {\enquote {\bibinfo {title} {Electronic structure of
  {Nd$_2$Fe$_{14}$B}},}\ }\href {\doibase 10.1063/1.338671} {\bibfield
  {journal} {\bibinfo  {journal} {J. Appl. Phys.}\ }\textbf {\bibinfo {volume}
  {61}},\ \bibinfo {pages} {3718--3720} (\bibinfo {year} {1987})}\BibitemShut
  {NoStop}%
\bibitem [{\citenamefont {Morin}\ and\ \citenamefont
  {Schmitt}(1990)}]{38morin1990quadrupolar}%
  \BibitemOpen
  \bibfield  {author} {\bibinfo {author} {\bibfnamefont {P.}~\bibnamefont
  {Morin}}\ and\ \bibinfo {author} {\bibfnamefont {D.}~\bibnamefont
  {Schmitt}},\ }\bibfield  {title} {\enquote {\bibinfo {title} {Quadrupolar
  interactions and magneto-elastic effects in rare earth intermetallic
  compounds},}\ }\href {\doibase 10.1016/S1574-9304(05)80061-6} {\bibfield
  {journal} {\bibinfo  {journal} {Handbook of Ferromagnetic Materials}\
  }\textbf {\bibinfo {volume} {5}},\ \bibinfo {pages} {1--132} (\bibinfo {year}
  {1990})}\BibitemShut {NoStop}%
\bibitem [{\citenamefont {Zickler}\ \emph {et~al.}(2015)\citenamefont
  {Zickler}, \citenamefont {Toson}, \citenamefont {Asali},\ and\ \citenamefont
  {Fidler}}]{new12zickler2015nanoanalytical}%
  \BibitemOpen
  \bibfield  {author} {\bibinfo {author} {\bibfnamefont {G.~A.}\ \bibnamefont
  {Zickler}}, \bibinfo {author} {\bibfnamefont {P.}~\bibnamefont {Toson}},
  \bibinfo {author} {\bibfnamefont {A.}~\bibnamefont {Asali}}, \ and\ \bibinfo
  {author} {\bibfnamefont {J.}~\bibnamefont {Fidler}},\ }\bibfield  {title}
  {\enquote {\bibinfo {title} {Nanoanalytical {TEM} studies and micromagnetic
  modelling of {Nd-Fe-B} magnets},}\ }\href {\doibase
  10.1016/j.phpro.2015.12.164} {\bibfield  {journal} {\bibinfo  {journal}
  {Phys. Procedia}\ }\textbf {\bibinfo {volume} {75}},\ \bibinfo {pages}
  {1442--1449} (\bibinfo {year} {2015})}\BibitemShut {NoStop}%
\bibitem [{\citenamefont {Yi}\ \emph {et~al.}(2016)\citenamefont {Yi},
  \citenamefont {Gutfleisch},\ and\ \citenamefont
  {Xu}}]{39yi2016micromagnetic}%
  \BibitemOpen
  \bibfield  {author} {\bibinfo {author} {\bibfnamefont {M.}~\bibnamefont
  {Yi}}, \bibinfo {author} {\bibfnamefont {O.}~\bibnamefont {Gutfleisch}}, \
  and\ \bibinfo {author} {\bibfnamefont {B.-X.}\ \bibnamefont {Xu}},\
  }\bibfield  {title} {\enquote {\bibinfo {title} {Micromagnetic simulations on
  the grain shape effect in {Nd}--{Fe}--{B} magnets},}\ }\href {\doibase
  10.1063/1.4958697} {\bibfield  {journal} {\bibinfo  {journal} {J. Appl.
  Phys.}\ }\textbf {\bibinfo {volume} {120}},\ \bibinfo {pages} {033903}
  (\bibinfo {year} {2016})}\BibitemShut {NoStop}%
\bibitem [{\citenamefont {Erokhin}\ and\ \citenamefont
  {Berkov}(2017)}]{40erokhin2016optimization}%
  \BibitemOpen
  \bibfield  {author} {\bibinfo {author} {\bibfnamefont {S.}~\bibnamefont
  {Erokhin}}\ and\ \bibinfo {author} {\bibfnamefont {D.}~\bibnamefont
  {Berkov}},\ }\bibfield  {title} {\enquote {\bibinfo {title} {Optimization of
  nanocomposite materials for permanent magnets: Micromagnetic simulations of
  the effects of intergrain exchange and the shapes of hard grains},}\ }\href
  {\doibase 10.1103/PhysRevApplied.7.014011} {\bibfield  {journal} {\bibinfo
  {journal} {Phys. Rev. Appl.}\ }\textbf {\bibinfo {volume} {7}},\ \bibinfo
  {pages} {014011} (\bibinfo {year} {2017})}\BibitemShut {NoStop}%
\bibitem [{\citenamefont {Bance}\ \emph
  {et~al.}(2014{\natexlab{b}})\citenamefont {Bance}, \citenamefont
  {Fischbacher}, \citenamefont {Schrefl}, \citenamefont {Zins}, \citenamefont
  {Rieger},\ and\ \citenamefont {Cassignol}}]{41bance2014micromagnetics}%
  \BibitemOpen
  \bibfield  {author} {\bibinfo {author} {\bibfnamefont {S.}~\bibnamefont
  {Bance}}, \bibinfo {author} {\bibfnamefont {J.}~\bibnamefont {Fischbacher}},
  \bibinfo {author} {\bibfnamefont {T.}~\bibnamefont {Schrefl}}, \bibinfo
  {author} {\bibfnamefont {I.}~\bibnamefont {Zins}}, \bibinfo {author}
  {\bibfnamefont {G.}~\bibnamefont {Rieger}}, \ and\ \bibinfo {author}
  {\bibfnamefont {C.}~\bibnamefont {Cassignol}},\ }\bibfield  {title} {\enquote
  {\bibinfo {title} {Micromagnetics of shape anisotropy based permanent
  magnets},}\ }\href {\doibase 10.1016/j.jmmm.2014.03.070} {\bibfield
  {journal} {\bibinfo  {journal} {J. Magn. Magn. Mater.}\ }\textbf {\bibinfo
  {volume} {363}},\ \bibinfo {pages} {121--124} (\bibinfo {year}
  {2014}{\natexlab{b}})}\BibitemShut {NoStop}%
\bibitem [{\citenamefont {Bance}\ \emph
  {et~al.}(2014{\natexlab{c}})\citenamefont {Bance}, \citenamefont {Seebacher},
  \citenamefont {Schrefl}, \citenamefont {Exl}, \citenamefont {Winklhofer},
  \citenamefont {Hrkac}, \citenamefont {Zimanyi}, \citenamefont {Shoji},
  \citenamefont {Yano}, \citenamefont {Sakuma}, \citenamefont {Ito},
  \citenamefont {Kato},\ and\ \citenamefont {Manabe}}]{42bance2014grain}%
  \BibitemOpen
  \bibfield  {author} {\bibinfo {author} {\bibfnamefont {S.}~\bibnamefont
  {Bance}}, \bibinfo {author} {\bibfnamefont {B.}~\bibnamefont {Seebacher}},
  \bibinfo {author} {\bibfnamefont {T.}~\bibnamefont {Schrefl}}, \bibinfo
  {author} {\bibfnamefont {L.}~\bibnamefont {Exl}}, \bibinfo {author}
  {\bibfnamefont {M.}~\bibnamefont {Winklhofer}}, \bibinfo {author}
  {\bibfnamefont {G.}~\bibnamefont {Hrkac}}, \bibinfo {author} {\bibfnamefont
  {G.}~\bibnamefont {Zimanyi}}, \bibinfo {author} {\bibfnamefont
  {T.}~\bibnamefont {Shoji}}, \bibinfo {author} {\bibfnamefont
  {M.}~\bibnamefont {Yano}}, \bibinfo {author} {\bibfnamefont {N.}~\bibnamefont
  {Sakuma}}, \bibinfo {author} {\bibfnamefont {M.}~\bibnamefont {Ito}},
  \bibinfo {author} {\bibfnamefont {A.}~\bibnamefont {Kato}}, \ and\ \bibinfo
  {author} {\bibfnamefont {A.}~\bibnamefont {Manabe}},\ }\bibfield  {title}
  {\enquote {\bibinfo {title} {Grain-size dependent demagnetizing factors in
  permanent magnets},}\ }\href {\doibase 10.1063/1.4904854} {\bibfield
  {journal} {\bibinfo  {journal} {J. Appl. Phys.}\ }\textbf {\bibinfo {volume}
  {116}},\ \bibinfo {pages} {233903} (\bibinfo {year}
  {2014}{\natexlab{c}})}\BibitemShut {NoStop}%
\bibitem [{\citenamefont {Schrefl}\ \emph {et~al.}(1994)\citenamefont
  {Schrefl}, \citenamefont {Fidler},\ and\ \citenamefont
  {Kronm{\"u}ller}}]{43schrefl1994nucleation}%
  \BibitemOpen
  \bibfield  {author} {\bibinfo {author} {\bibfnamefont {T.}~\bibnamefont
  {Schrefl}}, \bibinfo {author} {\bibfnamefont {J.}~\bibnamefont {Fidler}}, \
  and\ \bibinfo {author} {\bibfnamefont {H.}~\bibnamefont {Kronm{\"u}ller}},\
  }\bibfield  {title} {\enquote {\bibinfo {title} {Nucleation fields of hard
  magnetic particles in 2{D} and 3{D} micromagnetic calculations},}\ }\href
  {\doibase 10.1016/0304-8853(94)90395-6} {\bibfield  {journal} {\bibinfo
  {journal} {J. Magn. Magn. Mater.}\ }\textbf {\bibinfo {volume} {138}},\
  \bibinfo {pages} {15--30} (\bibinfo {year} {1994})}\BibitemShut {NoStop}%
\end{thebibliography}%

\end{document}